\documentclass[10pt,a4paper]{article}
\pdfoutput=1
\usepackage{jcappub}
\usepackage[english]{babel} 
\usepackage{commath}
\usepackage{chngcntr}

\def\omo{\Omega_{\mathrm{m},0}}
\def\om{\Omega_{\mathrm{m}}}
\def\m22{\mu_2^2}
\def\ak{\alpha_K}
\def\mps{M_{\rm pl}^2}
\def\gn{G_{\rm N}}
\def\muff{\mu_{\rm ff}}
\def\musc{\mu_{\rm sc}}
\def\gsp{\gamma}
\def\sig{\sigma_8} 
\def\fs{f\sigma_8}
\def\so{\sigma_{8,0}}
\def\gdot{\dot{G}_{\rm N}}
\def\pij{p_{ij}}
\def\p10{p_{10}}
\def\mod{H3 }
\def\lcdm{$\Lambda$CDM }
\newcommand{\vv}{\vskip2mm}
\def\fps{$f+\sigma_8$ } 
\def\fspfps{$f\sigma_8+f+\sigma_8$ } 
\graphicspath{{figures/}}
\usepackage[scale=0.8]{geometry}

\begin{document}

\hypersetup{pageanchor=false} 
\title{Optimising growth of structure constraints on modified gravity}

\author[a,b]{Louis Perenon}
\author[c]{, Julien Bel}
\author[a,d]{, Roy Maartens}
\author[b]{, Alvaro de la Cruz-Dombriz}

\affiliation[a]{Department of Physics \& Astronomy, University of the Western Cape, Cape Town 7535, South Africa}

\affiliation[b]{Cosmology and Gravity Group, Department of Mathematics and Applied Mathematics, University of Cape Town, Rondebosch 7701, Cape Town, South Africa}

\affiliation[c]{Aix Marseille Univ, Universit\'e de Toulon, CNRS, CPT, Marseille, France}

\affiliation[d]{Institute of Cosmology \& Gravitation, University of Portsmouth, Portsmouth PO1 3FX, UK}

\emailAdd{lperenon@uwc.ac.za}

\abstract{We use growth of structure data to constrain the effective field theory of dark energy. Considering as case study Horndeski theories with the speed of gravitational waves equal to that of light, we show how constraints on the free parameters and the large-scale structure phenomenological functions can be improved by two ingredients: firstly by complementing the set of redshift-space distortions data with the three recent measurements of the growth rate $f$ and the amplitude of matter fluctuations $\sig$ from the VIPERS and SDSS collaborations; secondly by applying a local Solar System bound on the variation of the Newton constant. This analysis allows us to conclude that: $i)$ despite firmly restricting the predictions of weaker gravity, the inclusion of the Solar System bound does not prevent suppressed growth relative to the standard model $\Lambda$CDM at low redshifts; $ii)$ the same bound in conjunction with the growth of structure data strongly restricts the redshift evolution of the gravitational slip parameter to be close to unity and the present value is constrained to one at the $10^{-3}$ level; $iii)$ the growth of structure data favours a fifth force contribution to the effective gravitational coupling at low redshifts and at more than two sigma at present time.}


\maketitle

\section{Introduction}

The number of dark energy models and modified gravity theories has considerably risen since the discovery of cosmic acceleration \cite{Perlmutter:1998np,Riess:1998cb}. The creation of common theoretical formulations to test the observational viability of various theoretical proposals on the same grounds, and the gathering of ever more precise data to constrain the growing degrees freedom in such proposals are thus crucial today. On the one hand, the effective field theory of dark energy \cite{Gubitosi:2012hu,Bloomfield:2012ff} (EFT of DE), through its unitary gauge construction of a gravitational action, renders the linear description of virtually all theories containing an extra scalar degree of freedom added to General Relativity (GR) under the same framework. A large class of theories are the so-called Horndeski theories \cite{Horndeski:1974wa}, or Generalised Galileons \cite{Deffayet:2009mn}, being the most general four dimensional scalar-tensor theories giving rise directly to field equations of motion at most second order. On the other hand, future ground based telescopes such as SKA\footnote{\url{https://www.skatelescope.org/}} \cite{Bacon:2018dui}, DESI\footnote{\url{https://www.desi.lbl.gov/}} \cite{Aghamousa:2016zmz,Aghamousa:2016sne}, LSST\footnote{\url{https://www.lsst.org/}} \cite{Mandelbaum:2018ouv} and satellites such as Euclid\footnote{\url{http://www.euclid-ec.org/}} \cite{Laureijs:2011gra} forecast great improvements on the observation of the large-scale structure of the Universe. Notably, measurements of the growth of structures at sub-percent level are expected \cite{Amendola:2012ys,Bull:2018lat,Sprenger:2018tdb} while our observational knowledge will also greatly benefit from the addition of up-coming Cosmic Microwave Background (CMB) Stage 4 experiments \cite{Abazajian:2016yjj}.

In this respect, a great effort is conducted in preparing the statistical tools to constrain any departures from the \lcdm Concordance Model more efficiently in the future. A powerful tool to complement the constraining power of data over Horndeski theories is the use of phenomenological and theoretical viability conditions \cite{Piazza:2013pua,Raveri:2014cka,delaCruz-Dombriz:2015tye,Peirone:2017lgi,Raveri:2017qvt,Peirone:2017ywi,Espejo:2018hxa,Frusciante:2018vht,Noller:2018eht}. Requiring for instance that a model must be stable from the Hamiltonian point of view and produce propagation speeds of perturbations smaller than or equal to that light significantly improves CMB constraints \cite{Salvatelli:2016mgy}. Recently, the merging of two neutron stars measured by the LIGO/Virgo collaboration with its electromagnetic counterpart by the FERMI satellite \cite{TheLIGOScientific:2017qsa,Monitor:2017mdv} has given rise to the era of multi-messenger cosmology \cite{Ezquiaga:2018btd} and can lead, in certain cases, to stringent bounds on modified gravity \cite{Lombriser:2015sxa,Lombriser:2016yzn,Creminelli:2017sry,Ezquiaga:2017ekz,Sakstein:2017xjx,Langlois:2017dyl,Boran:2017rdn,Casalino:2018tcd,Casalino:2018wnc}. See \cite{Kase:2018aps} for a recent review. 

Unfortunately astrophysical scales offer only a small window for constraining modified gravity as they correspond to a screened environment and on these scales GR has been tested to an incredible precision \cite{Uzan2011}. Any modified gravity theory in agreement with such observational knowledge should in principle yield predictions virtually equal to those of GR thanks to the screening mechanism, rendering the possibility to probe modifications of gravity almost in vain. In the context of Horndeski theories, this picture is only partially true given un-screenable effects already present at linear level \cite{Perenon:2015sla} where the Vainstein screening mechanism \cite{Vainshtein:1972sx} is ``pierced'' by the gradient of the scalar field  \cite{Jimenez:2015bwa}, $i.e.$ the latter does not vanish necessarily in the screened region. This, for instance, renders the application of the stringent bounds on the variation of the Newton constant necessary. Galileon models \cite{Nicolis:2008in} have also been severely constrained by Solar System  tests \cite{Barreira:2014jha}. Beyond the Horndeski landscape, GLPV theories \cite{Gleyzes:2014dya} display an additional breaking of the Vainstein mechanism \cite{Kobayashi:2014ida}. This gives the possibility to constrain modifications of gravity with massive and compact objects \cite{Saito:2015fza,Sakstein:2015zoa,Sakstein:2016ggl,Koyama:2015oma,Sakstein:2016oel,Sakstein:2017xjx}. In parallel, modifications of gravity effects can be assessed from the difference of exterior space-time solutions for neutron and quarks stars which differ from GR in modified gravity theories \cite{delaCruz-Dombriz:2014zaa,Resco:2016upv,Astashenok:2017dpo}.

On cosmological scales, the growth of structures is a crucial probe to constrain low redshift departures from standard gravity and until now, the standard observable has been the growth function $\fs$. Recently the VIPERS collaboration released the first two disentangled measurements from redshiftspace distortions of the growth rate, $f$, and the root mean squared of matter fluctuations averaged over a sphere of $8$ h$^{-1}$Mpc $\sig$ \cite{delaTorre:2016rxm} thanks to the combination with galaxy-galaxy lensing. This was followed by the release of one measurement of $f$ and one of $\sig$ by the SDSS collaboration \cite{Shi:2017qpr}. The use of $f$ and $\sig$ separated opens a new window to constrain departures from the standard model in the perturbation sector. 

In the present paper we put forward the idea of optimising cosmological constraints further by considering, beyond viability requirements, on the one hand the inclusion of $f$ and $\sig$ separated data and on the other hand, a stringent bound on the variation of Newton's constant. To investigate the impact of these considerations on growth of structures constraints, we use the EFT of DE and focus on Horndeski theories with the speed of gravitational waves equal to the speed of light. Our goal is to concentrate on growth of structure constraints on modified gravity while the inclusion of other cosmological probes is left for future work. We refer the reader to the literature for more complete current constraints \cite{Bellini:2015xja,Bellomo:2016xhl,Kreisch:2017uet,Noller:2018wyv,SpurioMancini:2019rxy} and forecasts \cite{Alonso:2016suf,Mancini:2018qtb,Reischke:2018ooh,Abazajian:2016yjj,Frusciante:2018jzw,Gleyzes:2015rua} on Horndeski theories in terms of cosmological probes. We also derive and discuss the constraints on large-scale structure (LSS) phenomenological functions key for characterising departures from GR \cite{2010PhRvD..81j4023P}, $e.g.$ the effective gravitational coupling $\mu$, the gravitational slip parameter $\gsp$ and the light deflection parameter $\Sigma$.

\vv
This paper is structured as follows. We first present how the EFT of DE is parametrised to explore Horndeski theories in Section 2. Then in Section 3 we discuss the improvements obtained from considering the VIPERS and SDSS measurements of $f$ and $\sig$ thoroughly while Section 4 is dedicated to the application of a strong bound on the variation of the Newton constant. We conclude in Section 5.

\section{Parametrising modified gravity}

The theories encapsulated in the EFT of DE formulation are numerous and go now beyond the linear Horndeski paradigm (see for instance \cite{Gleyzes:2014qga,Gleyzes:2015pma,DAmico:2016ntq,Langlois:2017mxy,Lagos:2017hdr,Cusin:2017mzw}). However, to be able to gauge more precisely the improvements that growth of structure constraints on modified gravity can undergo, we restrict our analysis to models of the Horndeski class. Furthermore, since the recent major discovery in physics, the first measurement of gravitational waves \cite{Abbott:2016blz}, achieved by the LIGO/Virgo collaboration, several events have been registered. In particular, the merging of two neutron stars detected with its electromagnetic counterpart by the FERMI satellite \cite{TheLIGOScientific:2017qsa,Monitor:2017mdv} has led to implications of paramount importance. One of them is the speed of gravitational waves now constrained to be extremely close to that of light, at the $10^{-15}$ level, at low redshifts. The implications of this stringent bound on modified gravity are yet to be fully assessed and one must be careful of concluding that it enforces theories to have $c_T=1$ at all times. For instance, viable quartic and quintic Lagrangians of Horndeski theories, $i.e.$ the Lagrangians responsible for a varying $c_T$, were rapidly concluded to be drastically reduced. It was however recently shown that some non-trivial functional forms can be ``rescued" \cite{Copeland:2018yuh}. Furthermore, from a more pragmatic standpoint, this new bound on $c_T$ must be assessed in time and scale. On the one hand, this bound was obtained at low redshifts $z \lesssim 0.01$, rendering it relevant for late time cosmic acceleration but not necessarily at earlier times. In fact, $c_T$ could vary during the evolution of the universe and reach unity at present time without any fine-tuning \cite{Kennedy:2018gtx}. On the other hand, the energy scale associated to the LIGO/Virgo detection differs by many orders of magnitude from that of cosmic acceleration \cite{Battye:2018ssx}. In fact, it stands very close to the cut-off scale of many dark energy models and Horndeski theories \cite{deRham:2018red}. It has been shown in \cite{deRham:2018red} how upon UV completion an anomalous speed of gravitational waves can be brought back to that of light for the frequencies observed by LIGO/Virgo. Nevertheless in this paper, we will adopt the restrictive approach and set $c_T=1$ for it serves our goal of remaining in a simplified setup to focus on the optimisation of growth of structure constraints. We emphasise this to be a choice not a necessity.

\subsection{Coupling functions and background evolution}\label{sec:param}

Following the convention introduced in \cite{Piazza:2013pua}, a useful rescaling of \cite{Gubitosi:2012hu}, the unitary gauge gravitational action describing linear perturbations in Horndeski theories with $c_T=1$ within the EFT of DE can be cast into

\begin{equation}\label{action}
S_g \ = \  \int \! {\rm d}^4x \sqrt{-g} \, \frac{M^2(t)}{2} \, \left[R \, -\,  2 \lambda(t) \, - \, 2 \mathcal{C}(t) g^{00} \,-\,\mu_2^2(t) (\delta g^{00})^2 \, -\, \mu_3(t) \, \delta K \delta g^{00} \right] \;, 
\end{equation}
where $t$ is cosmic time, $g$ is the determinant of the metric, $\delta g^{00}= 1+g^{00}$ and $K$ is the trace of the extrinsic curvature tensor. The contributions of the scalar field to the background energy momentum tensor, $\lambda(t)$ and $\mathcal{C}(t)$, are not free functions in Horndeski theories but are expressed as 
\begin{align}\label{candlambda}
\mathcal{C} & =  \dfrac12 \left( H\mu_1-\dot\mu_1-\mu_1^2 \right) - \dot H - \frac{3}{2}\frac{\mps}{M^2} H^2 \om  \ , \\
\lambda     & =  \dfrac12 \left( 5H\mu_1+\dot\mu_1+\mu_1^2 \right) + \dot H +3H^2-\frac{3}{2}\frac{\mps}{M^2} H^2 \om  \ ,
\end{align}
where $\mps$ is the Planck mass. From this configuration, the set of EFT of DE coupling functions to supply in order to model the time evolution of the linear perturbations amounts to $\lbrace \mu_1(t)={\rm d}\ln M^2(t)/{\rm d}\ln t,\,\mu^2_2(t),\,\mu_3(t)\rbrace$ in addition to providing the Hubble rate $H(t)$. The Brans-Dicke \cite{Brans:1961sx} coupling function or running Planck mass corresponds to $\mu_1(t)$ which tracks the time variation of the bare Planck mass $M^2(t)$. The remaining two coupling functions arise from considering up to the Cubic Galileon \cite{Nicolis:2008in} part of Horndeski theories. We will therefore denote this restriction of Horndeski theories with $c_T=1$ by \mod following the terminology of \cite{Perenon:2015sla}. 

Growth of structure measurements are generically computed in a \lcdm fiducial background. We shall therefore adopt the same convention for the sake of comparison. Using the EFT of DE framework, one is granted the possibility to do so straightforwardly by fixing the Hubble rate $H(t)$ to ones choice. Choosing that of a spatially flat-\lcdm model, we fix the redshift evolution of the Hubble rate thus
\begin{equation} \label{h}
\frac{H^2(z)}{H_0^2} = \omo (1+z)^3 + 1-\omo \; ,
\end{equation}
where $\omo$ is the fractional matter density parameter evaluated today and $H_0$ the Hubble constant. According to eq. \eqref{h}, one can obtain the redshift evolution of the reduced matter density parameter $\om(z)$ of the fiducial background as 
\begin{equation}
\label{xdef}
\om(z)\ = \ \frac{\omo}{\omo + (1- \omo) (1+ z)^{-3}}\; .
\end{equation}
The focus of this paper being on data sets of the growth of structures and on models giving rise to modifications of gravity up to the matter domination epoch, it allows us to neglect radiation in the previous expression. For this simplified choice of background the only free parameter needed to be constrained by data is therefore $\omo$. 

The time variation of the EFT couplings cannot be derived from the underlying theory, hence highlighting the functional freedom of Horndeski theories. One can adopt phenomenological parametrisations so as to conduct a likelihood analysis. A common and simple parametrisation of the couplings used in the literature corresponds to modelling their evolution by a constant times the effective dark energy density parameter $\Omega_\mathrm{de}(z) = 1- \Omega_m(z)$, see for example \cite{Gleyzes:2015rua,Bellini:2015xja,Huang:2015srv,Alonso:2016suf}. Allowing one free parameter per coupling will be the least CPU greedy for statistical endeavours; however, such a parametrisation scheme can be too crude and is likely to yield poor fits \cite{Linder:2015rcz,Linder:2016wqw,Gleyzes:2017kpi}. Therefore, one is bound to find the best middle ground between the generality and efficiency of the parametrisations and reasonable CPU time\footnote{Note that, when possible, more involved parametrisations can and have been explored, see for example the use of Pad\'e functions in \cite{Peirone:2017ywi}, and the analysis comparing several parametrisations conducted in \cite{Gleyzes:2017kpi}.}. In \cite{Perenon:2015sla,Perenon:2016blf}, a polynomial expansion in $(\om(z) -\omo)$ up to order two, hence, three free parameters per coupling function, was adopted to capture all the rich phenomenology of Horndeski theories. In light of the above, we split the difference and consider up to two free parameters $\pij$ per coupling: 
\begin{align}
\label{parammu1}\frac{\mu_1}{H}(z)      \ &= \frac{1-\om(z)}{1-\omo} \left[p_{10}+p_{11}\left(\om(z)-\omo\right)\right]\;,\\[1mm]
\frac{\mu^2_2}{H^2}(z)  \ &= \frac{1-\om(z)}{1-\omo} \left[p_{20}+p_{21}\left(\om(z)-\omo\right)\right]\;,\\[1mm]
\frac{\mu_3}{H}(z)      \ &= \frac{1-\om(z)}{1-\omo} \left[p_{30}+p_{31}\left(\om(z)-\omo\right)\right]\;.
\end{align}
The couplings above have a pre-factor $(1-\om(z))$ so as to make sure they vanish in the past and they are normalised by $(1- \omo)$ to obtain the constraints on the dimensionless coupling at present time to be directly that of the first free parameter $p_{i0}$, $e.g.$ $\mu_3(z=0)/H_0=p_{30}$. Lastly, to discuss the implications of growth of structure constraints, we put in perspective two levels of generality of this parametrisation. We consider a one and two dimensional like parametrisation of each coupling where the nonzero free parameters are
\begin{align}
\label{eq:param_order0}{\rm 1D}\,:\; & \lbrace \omo,\,\so,\,p_{10},\,p_{20},\,p_{30}, \rbrace \; ,\\[1mm]
\label{eq:param_order1}{\rm 2D}\,:\; & \lbrace \omo,\,\so,\,p_{10},\,p_{11},\,p_{20},\,p_{21},\,p_{30},\,p_{31} \rbrace \; ,
\end{align} 
and $\so$ is the root mean square of matter fluctuations evaluated at present time. This is the set of parameters we will constrain with growth of structure data in the likelihood analysis. 

\subsection{Characterising the large-scale structure}\label{sec:characLSS}

Given the landscape of our analysis, we adopt two simplifying assumptions to derive characteristic phenomenological functions of the LSS sensitive to the modifications of gravity induced by the EFT couplings: we neglect any scale dependence and we consider the quasi-static approximation (QSA), as in \cite{Piazza:2013pua,Perenon:2015sla,Perenon:2016blf}. On the one hand, the mass of the scalar extra degree of freedom contained in Horndeski theories must be of order Hubble to produce cosmic acceleration. This implies that any scale dependence would be seen around Hubble scales, and such a scale  dependence remains irrelevant for the growth of structure data that we consider. On the other hand, the QSA has been shown to provide a faithful description as long as the scales considered remain within the sound horizon of the scalar degree of freedom \cite{Sawicki:2015zya,Frusciante:2018jzw}. To ensure that the models we consider respect this criterion, following \cite{Sawicki:2015zya}, we impose a lower bound on the speed of sound of dark energy perturbations for future surveys (see Section \ref{sec:priors}). Furthermore, the QSA was shown numerically to hold strongly at small and intermediate scales, but to break down at present time for $k \lesssim 0.001\; {\rm h/Mpc}$ \cite{Peirone:2017ywi}, for a large and general sample of viable Horndeski models. Hence the breakdown of the QSA for the models we consider remains safely outside of the range probed by the growth of structure data we use.
\vv
The quantity characterising the modification of the dynamics of gravitating bodies with respect to GR is the effective gravitational coupling, or effective Newton constant, $\mu$. Considering the flat Newtonian gauge ${\rm d}s^2 = - (1+ 2 \Phi) {\rm d}t^2 + a^2 (1 - 2 \Psi) \delta_{i j} {\rm d} x^i {\rm d} x^j$, the latter is defined, following \cite{Gleyzes:2013ooa}, by the modified Poisson equation $k^2 / a^2\, \Phi =-\frac{3}{2} H^2\, \om \,\mu \,\delta_m$, where $k$ is the comoving Fourier mode wave number, $a$ is the scale factor of the universe and $\delta_m$ is the linear density perturbation of matter. In \cite{Perenon:2015sla}, the effective gravitational coupling was shown to be the result of two contributions,
\begin{equation}\label{eq:mu}
\mu = \musc \left( 1 +\muff\right)\;,
\end{equation} 
translating into two phenomenologically different signatures of modified gravity. In a screened environment, $i.e.$ an environment where the extra scalar field is decoupled from the matter fields, the dynamics of matter fields remain affected by un-screenable modifications of gravity. Two ways of obtaining $\musc$ have been developed in the literature:  take the super-Compton limit of the effective gravitational coupling  \cite{Pogosian:2016pwr} or work with the action \eqref{action} written in the Newtonian gauge \cite{Perenon:2015sla}. With the second method it is straightforward to identify the gravitational coupling which remains when the extra scalar field is decoupled from gravity. This requires cancellation of the self-interaction term of the latter and its interaction terms with the metric fields -- the gravitational coupling that remains is $G_\mathrm{sc} = 1/(8\pi M^2)$. Furthermore, the fact that we live in such a screened environment imposes this screened gravitational coupling measured today $G_\mathrm{sc}(z=0) \equiv \gn$, or equivalently, the Planck mass to be defined as the bare Planck mass $M^2$ today, $i.e.$ $\mps \equiv M^2(z=0)$. Therefore, by normalising the screened gravitational coupling accordingly one obtains
\begin{equation}\label{eq:musc}
\musc = \frac{M^2(z=0)}{M^2(z)}\;,
\end{equation}
which implies $\musc(z=0)=1$. The second contribution in \eqref{eq:mu}, $\muff$, corresponds to the fifth force mediated by the scalar field. This is indeed the screenable part of the total modified gravity effect induced by the extra degree of freedom. Knowing $\mu$ and $\musc$, it is straightforward to deduce  that this contribution yields
\begin{equation}\label{eq:muff}
\muff = \frac{(\mu_1+\mu_3)^2}{2B}\;.
\end{equation}
One can clearly observe the latter to be positive since $B$ is the gradient stability condition and it must be positive for a healthy spin 0 field (see Section \ref{sec:priors} for its expression). Further details on the discussion in this paragraph and the specific analytical derivations can be found in \cite{Perenon:2015sla}.

\vv
Beyond understanding the nature of the modifications of gravitating bodies, additional LSS phenomenological functions prove useful to quantify and detect observational departures from standard gravity \cite{Song:2010fg,Simpson:2012ra,Ade:2015rim,Aghanim:2018eyx}. Therefore, we will take a close look at the constraints growth of structures yield on the correlations of $\mu$, the gravitational slip parameter $\gsp$ and the light deflection parameter $\Sigma$. The three LSS functions are linked by the relation
\begin{equation} \label{Sigma}
\Sigma=\mu\,\frac{1+\gsp}{2} \; .
\end{equation}
The gravitational slip parameter, $i.e.$ the ratio between the gravitational potentials $\gsp=\Psi/\Phi$, is given as a function of the couplings
\begin{equation} \label{gsp}
\gsp=1- \dfrac{\mu_1 (\mu_1+ \mu_3 )}{\left(H+\mu_1\right)\left(\mu_1+\mu_3\right)-\dot\mu_1+\dot\mu_3 - 2\dot H - 3(\mps/M^2) H^2 \om} \;.
\end{equation}
While $\gsp$ is of difficult observability, the sensitivity of the light deflection parameter to the Weyl potential, $k^2 / a^2\, (\Phi+\Psi) =-3 H^2 \om\, \Sigma\,\delta_m$, makes it a more powerful candidate for modified gravity constraints from weak lensing surveys. Note that the distinction between screened and fifth force contributions can be derived for $\gsp$ and $\Sigma$ also \cite{Perenon:2015sla}.

\vv
The total effective gravitational coupling $\mu$ plays a significant role in the growth of structures. Next to $H(t)$ and $\om(t)$, it acts as a regulator of the evolution of the linear matter density perturbations $\delta_m$ where the Newton constant $\gn$ must be promoted to $\mu$ \cite{Piazza:2013pua} in $\ddot{\delta}_m(t) +2H(t)\dot{\delta}_m(t)- (3/2)\,H^2(t)\,\om(t) \,\mu(t) \, \delta_m(t)=0$. The dot corresponds to a derivative with respect to cosmic time. Having neglected the scale dependence in $\mu$, we can write $\delta_m (k,z)=\delta_k\, D_+(z)$ and consider only the linear growing mode $D_+$ of matter fluctuations. A little algebra suffices to show that the latter, given the Hubble rate of eq. \eqref{h}, implies
\begin{equation}\label{eq:evolDplus}
D''_+ +\left(\frac{H'}{H}-\frac{1}{1+z}\right)\, D'_+  -\frac{3}{2}\frac{\om }{\left(1+z\right)^2} \,\mu\, D_+=0 \; ,
\end{equation}
where the prime corresponds to a derivative with respect to the redshift. The initial conditions for $D_+$ should be set according to the problem at hand. Since it is not possible to follow a single perturbation across cosmic time, we use a statistical description, based on the root mean square (rms) of linear matter fluctuations in spheres of radius $R=8h^{-1}$Mpc, defined as $\sig^2(z) \equiv \langle\delta_R^2(z)\rangle$, where $\delta_R(z) = D_+(z)/D_+(z_\mathrm{ini})\delta_R(z_\mathrm{ini})$ for a given initial redshift $z_\mathrm{ini}$. The value of $\sig (z)$ must be allowed to freely evolve until today according to the modified gravity effects induced by the EFT couplings, while its initial past value should be in agreement with CMB constraints. The technicalities of this procedure are explained in detailed in Section \ref{sec:priors}. Nevertheless, to do so, $\sig (z)$ must be normalised to its observed value at initial time
\begin{equation}\label{eq:normsig}
\sig (z)= \sig (z_\mathrm{ini}) \frac{D_{+}(z)}{D_{+} (z_\mathrm{ini})} \; ,
\end{equation}
where we fix $z_\mathrm{ini}$ large enough so as to make sure $f$ and $\sig$ have reached the attractor solution early enough for the application of observational bounds. From this normalisation, one can simply set the initial condition $D_+(z_\mathrm{ini})=1$. Furthermore, since the initial value $\sig$ will be made in agreement with CMB observations deep in matter domination, hence not far from \lcdm predictions, one can apply the usual prescription $\dot{D}_+(t_\mathrm{ini})\sim H(t_\mathrm{ini}) D_+(t_\mathrm{ini})$. The evolution of the growth rate, $f={\rm d}\ln D_{+}/{\rm d} \ln a$, can also be derived from eq. \eqref{eq:evolDplus} which yields,
\begin{equation}\label{eq:evolf}
\left(1+z\right) f'-f^2+\left[ (1+z)\frac{H'}{H}-2 \right] f+\frac{3}{2}\, \om \,\mu=0\; ,
\end{equation} 
Following the same prescription as for $D_+(z_\mathrm{ini})$, we set $f(z_\mathrm{ini})=1$. The set of observables $\lbrace  (\fs) (z),\,f(z),\,\sig (z)  \rbrace$ is the one we have here at our disposal to characterise and constrain modified gravity effects with the growth of structures.

\subsection{Theoretical viability and CMB normalisation}\label{sec:priors}

Beyond having no prior information on the time behaviour of the EFT couplings and requiring phenomenological parametrisations, the coupled behaviour of the couplings is ultimately restricted by viability conditions. To ensure a model is respectively free of ghost and gradient instabilities, the conditions
\begin{align} \label{Acond}
A \ &= \ 2\mathcal{C} + 4 \mu_2^2 + \frac32 (\mu_1-\mu_3)^2 > \ 0\ , \\ \label{Bcond}
B \ &= \ 2\mathcal{C} + \dot\mu_3 + H \mu_3 + \frac12(3\mu_1^2 -\mu_3^2) \geqslant \ 0\  , 
\end{align}
must be met at all times. The speed of sound of dark energy perturbations is also required to be large enough for the QSA to be valid in the context of future surveys \cite{Sawicki:2015zya} $c_s^2=B/A\ge 0.1$. A particularity arising in the function $A$ has to be mentioned at this point. At the QSA level, $A$ is the sole expression where the coupling function $\m22$ appears, in other words, it does not appear in any of the expressions of the LSS phenomenological functions. Therefore, one would be tempted to set it to zero; however this coupling only acting through stability conditions can have important effects on the evolution of LSS observables by allowing more models to be stable. Indeed, being only a back-reaction, its behaviour is expected to be constrained poorly by data \cite{Bellini:2015xja,Kreisch:2017uet,Frusciante:2018jzw}. However, from its crucial role in scaling the stability conditions it must not be neglected in cosmological analysis \cite{Frusciante:2018jzw}. Lastly, placing an upper bound on $c_s$, such as $c_s\leqslant 1$, has been shown to optimise significantly the constraints on the EFT parameters \cite{Salvatelli:2016mgy}. However, as it was shown in \cite{Perenon:2016blf}, this upper bound becomes irrelevant once the coupling $\m22$ is not fixed. In fact, we have verified that our constraints are virtually unchanged with or without a $c_s\leqslant 1$ prior.

\vv
As mentioned previously, we do not allow the early time behaviour of the model to be fully free and a prior constraint is applied on the pair $\lbrace\omo,\, \sig (z_\mathrm{cmb}) \rbrace$ to avoid any conspicuous departures. We do so by using the Planck covariance matrix\footnote{\url{https://pla.esac.esa.int}} on these two parameters. The reasons leading to this choice are the following. We found growth of structure measurements to lack power in constraining $\omo$ hence adding prior knowledge is justified. Secondly, one must also guarantee that the $\so$ which an EFT model produces is in agreement with CMB constraints. To do so we use the rescaling method of \cite{Perenon:2015sla,Perenon:2016blf}. The Planck covariance matrix is a constraint at present time, so that applying this bound directly on the present day value $\so$ of a modified gravity model would bias the analysis. This constraint is the result of extrapolating a CMB constraint at present time in a $\Lambda$CDM model, and would therefore force wrongly the modified gravity model to do so as well. To bypass this problem, one must rescale rather the value of $\sig (z_\mathrm{cmb})$ that is predicted by the Horndeski models, to today, according to the Planck \lcdm cosmology,
\begin{equation}\label{eq:sigrescale}
\sigma_{8,0}^*=\sigma_{8}(z_\mathrm{cmb}) \,\frac{D_+^{\Lambda\rm CDM}(z=0)}{D_+^{\Lambda\rm CDM}(z_\mathrm{cmb})} \; ,
\end{equation}
where $z_\mathrm{cmb}=1090$. The Planck covariance matrix $C^{-1}_{\rm Planck}$ can then be used appropriately to compute the $\chi^2$ of $(\omo,\, \so^*)$\footnote{Another possibility would have been to translate the Planck covariance matrix in a \lcdm model to $\sigma_{8}(z_{\rm ini})$.}, 
\begin{equation}
\chi^2_{(\omo,\, \so^*)}=(\omo-\omo^{\rm Planck},\sigma_{8,0}^*-\sigma_{8,0}^{\rm Planck})\;C^{-1}_{\rm Planck}\;(\omo-\omo^{\rm Planck},\sigma_{8,0}^*-\sigma_{8,0}^{\rm Planck})^{t} \; ,
\end{equation}
where we set $\omo^{\rm Planck}=0.315$, $\sigma_{8,0}^{\rm Planck}=0.829$ \cite{Ade:2015xua}. We thereby define the Planck \lcdm fiducial model as $\lbrace \omo^{\rm Planck},\,\sigma_{8,0}^{\rm Planck},\, \pij=0 \rbrace$. The $2\times2$ covariance matrix $C_{\rm Planck}$ includes cross correlations and is extracted from the latest publicly available data release, $i.e.$ the 2015 CMB spectra. We consider the TT, EE, BB, and TE likelihood in the low-multipole range and the TT likelihood in the high-multipole range (see \cite{Ade:2015xua} for a complete description). Note that by the definition of the initial conditions, one has $D_+^{\Lambda\rm CDM}(z_\mathrm{ini})=D_{+}(z_\mathrm{ini})$, therefore using eq. \eqref{eq:normsig}, eq. \eqref{eq:sigrescale} becomes
\begin{equation}\label{rescale3}
\sigma_{8,0}^*=\sigma_{8,0} \frac{D_+^{\Lambda\rm CDM}(z=0)}{D_{+}(z=0) } \,.
\end{equation}
In summary, $\so$ is the value generated at each step of the likelihood analysis and $\so^*$ is its corresponding value required to be in agreement with CMB constraints through the amplitude of scalar modes ($A_s$). 

\section{Improving constraints with current data}\label{sec:currentcons}

\begin{table}
\small
\centering
\begin{tabular}{c c c c c c}
\hline
Dataset & $z$ & $\fs$ & $f$ & $\sigma_8$ & Ref. \\
\hline\hline
2MTF         & 0.001 & 0.505 $\pm$ 0.085 & & & \cite{Howlett:2017asq}\\
6dFGS+SNIa   & 0.02 & 0.428 $\pm$ 0.0465 & & & \cite{Huterer:2016uyq}\\
IRAS+SNIa    & 0.02 & 0.398 $\pm$ 0.065 & &  & \cite{2012ApJ...751L..30H,2012MNRAS.420..447T}\\ 
2MASS        & 0.02 & 0.314 $\pm$ 0.048 & & & \cite{2012ApJ...751L..30H,Davis:2010sw}\\ 
SDSS   & 0.10 & 0.376 $\pm$ 0.038 & 0.464 $\pm$ 0.040 & 0.769 $\pm$ 0.105 & \cite{Shi:2017qpr}\\ 
SDSS-MGS     & 0.15 & 0.490 $\pm$ 0.145 & & & \cite{Howlett:2014opa}\\ 
2dFGRS       & 0.17 & 0.510 $\pm$ 0.060 & & & \cite{Song:2008qt}\\ 
GAMA         & 0.18 & 0.360 $\pm$ 0.090 & & & \cite{Blake:2013nif}\\ 
GAMA         & 0.38 & 0.440 $\pm$ 0.060 & & &\cite{Blake:2013nif}\\ 
SDSS-LRG-200 & 0.25 & 0.3512 $\pm$ 0.0583 & & & \cite{2012MNRAS.420.2102S}\\ 
SDSS-LRG-200 & 0.37 & 0.4602 $\pm$ 0.0378 & & & \cite{2012MNRAS.420.2102S}\\ 
BOSS DR12    & 0.31 & 0.469 $\pm$ 0.098 & & &  \cite{Wang:2017wia}\\
BOSS DR12    & 0.36 & 0.474 $\pm$ 0.097 & & &  \cite{Wang:2017wia}\\
BOSS DR12    & 0.40 & 0.473 $\pm$ 0.086 & & &  \cite{Wang:2017wia}\\
BOSS DR12    & 0.44 & 0.481 $\pm$ 0.076 & & &  \cite{Wang:2017wia}\\
BOSS DR12    & 0.48 & 0.482 $\pm$ 0.067 & & &  \cite{Wang:2017wia}\\
BOSS DR12    & 0.52 & 0.488 $\pm$ 0.065 & & &  \cite{Wang:2017wia}\\
BOSS DR12    & 0.56 & 0.482 $\pm$ 0.067 & & &  \cite{Wang:2017wia}\\
BOSS DR12    & 0.59  & 0.481 $\pm$ 0.066 & & &  \cite{Wang:2017wia}\\
BOSS DR12    & 0.64  & 0.486 $\pm$ 0.070 & & &  \cite{Wang:2017wia}\\
WiggleZ      & 0.44  & 0.413 $\pm$ 0.080 & & & \cite{2012MNRAS.425..405B}\\ 
WiggleZ      & 0.60  & 0.390 $\pm$ 0.063 & & & \cite{2012MNRAS.425..405B}\\ 
WiggleZ      & 0.73  & 0.437 $\pm$ 0.072 & & & \cite{2012MNRAS.425..405B}\\ 
Vipers PDR-2 & 0.60  & 0.550 $\pm$ 0.120 & 0.93 $\pm$ 0.22 & 0.52 $\pm$ 0.06 & \cite{delaTorre:2016rxm,Pezzotta:2016gbo}\\ 
Vipers PDR-2 & 0.86  & 0.400 $\pm$ 0.110 & 0.99 $\pm$ 0.19 & 0.48 $\pm$ 0.04 & \cite{delaTorre:2016rxm,Pezzotta:2016gbo}\\ 
FastSound    & 1.40  & 0.482 $\pm$ 0.116 & & & \cite{Okumura:2015lvp}\\ 
SDSS-IV      & 0.978 & 0.379 $\pm$ 0.176 & & &  \cite{Zhao:2018jxv}\\
SDSS-IV      & 1.23  & 0.385 $\pm$ 0.099 & & &  \cite{Zhao:2018jxv}\\
SDSS-IV      & 1.526 & 0.342 $\pm$ 0.070 & & &  \cite{Zhao:2018jxv}\\
SDSS-IV      & 1.944 & 0.364 $\pm$ 0.106 & & &  \cite{Zhao:2018jxv}\\
\hline
\end{tabular} 
\caption{The growth of structure data used in the current analysis. The true uncertainty on the $\sig$ measurement of SDSS-veloc is $0.769_{-0.089}^{+0.121}$ but for simplicity we consider the symmetric uncertainty $0.769 \pm 0.105 $.}
\label{tab:data}
\end{table}
Having specified the theoretical background and observable characteristics, we now proceed with the constraints from the growth of structure data presented in Table \ref{tab:data}. The RSD data on the growth function $\fs$ are numerous and often display a tension with the standard model \cite{Kazantzidis:2018rnb,Kazantzidis:2018jtb}. For safety regarding the correlation between measurements, we restrict our analysis to the robust and independent collection compiled in \cite{Nesseris:2017vor} completed by the very low redshift release from \cite{Howlett:2017asq} and the higher redshift releases in \cite{Wang:2017wia} and \cite{Zhao:2018jxv}. Within this dataset, the measurements of WiggleZ \cite{2012MNRAS.425..405B} are correlated and so are those of SDSS-IV \cite{Zhao:2018jxv} and BOSS DR12 \cite{Wang:2017wia}. We therefore use their covariance matrix in our analysis, following \cite{Nesseris:2017vor,Sagredo:2018ahx}. The recent measurements of $f$ and $\sig$ from the VIPERS \cite{delaTorre:2016rxm} and SDSS collaboration \cite{Shi:2017qpr} also appear in the table. The use of galaxy-galaxy lensing was instrumental in providing a measurement of $f$ and $\sig$ separated. While RSD measurements suffer from a degeneracy in the combinations $\fs$ and $b_1\sig$, where $b_1$ is the linear biasing parameter, galaxy-galaxy lensing shows a degeneracy along $b_1\sigma_8^2$ and $b_1\sigma_8$ \cite{delaTorre:2016rxm}. Combining the two probes thereby allows one to break such degeneracies and enables one to produce a measurement of $f$ and $\sig$ out of $\fs$. We will compare constraints obtained from three data sets: {\small $i)$} the thirty RSD measurements of $\fs$, we call $\fs$; {\small $ii)$} The combination of the measurements of $f$ and three of $\sig$ from the VIPERS and SDSS collaborations, labelled as $f+\sig$; {\small $iii)$} The full combination of the twenty seven $\fs$, three $f$ and three $\sig$ measurements, labelled as $\fs+f+\sig$. For the latter case, the $\fs$ measurements which have their $f$ and $\sig$ counterpart are indeed not taken into account for the independence of the data. 

We explore the likelihood of the \mod model (see eq. \eqref{action}) using a Markov chain Monte Carlo (MCMC) procedure based on the Affine-Invariant Ensembler Sampler of \cite{ForemanMackey:2012ig}. The viability requirements, $i.e.$ the no ghost, no gradient instability and $c_s^2>0.1$ conditions, are imposed as hard priors. Any model not satisfying one of these conditions at any time in the past is automatically discarded. These viability priors and the Planck prior on $\lbrace\omo,\, \so^*\rbrace$ are imposed in all the runs.

\subsection{Optimising the constraints on the parameters}

\begin{figure}[!]
\begin{center}
\includegraphics[scale=0.4]{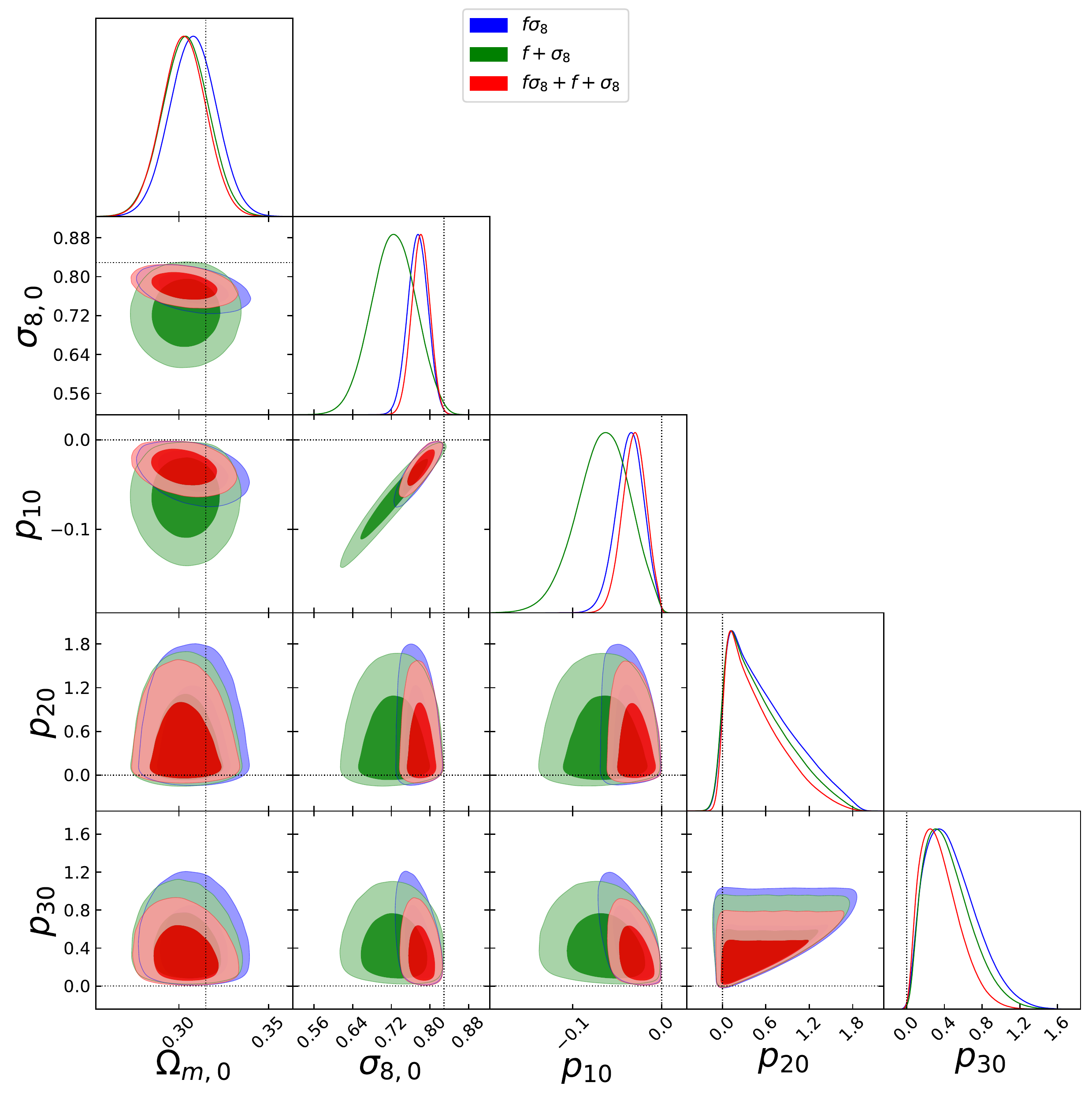}
   \caption{The constraints obtained from the data compiled in Table \ref{tab:data} on the free parameters of the \mod model \eqref{action} in the 1D parametrisation \eqref{eq:param_order0}. The constraints are derived for the three data sets as displayed in the label of the figure. The contours correspond to the posterior for the 2D-marginalised 68\% and 95\% confidence regions and the 1D marginalised posteriors are normalised to peak at one. The dotted lines signal the values in the standard model $\Lambda$CDM.}
   \label{fig:params}
\end{center}
\end{figure}

To evaluate and discuss the changes in the constraints obtained from the three data sets, we use the 1D parametrisation given by \eqref{eq:param_order0}. In this configuration, keeping always in mind that only $p_{10}$ characterises $\musc$, or equivalently $M^2$, hence the sole contribution to $\mu$ leading to weaker gravity, one striking feature to observe in Figure \ref{fig:params} is the favouring of suppressed growth with respect to \lcdm. Taking the example of the constraints obtained with $\fs$, the results yield a mean value of $p_{10}= -0.036_{-0.029}^{+0.030}$ (95\% c.l.) so as to induce a positive starting kick on the bare Planck mass which then increases with redshift. The negative value of the latter parameter is to be put in perspective with the mean value of $\so=0.775_{-0.041}^{+0.040}$  (95\% c.l.) and one can observe how these two parameters are naturally correlated while they exclude the Planck fiducial value at more than two sigma. We find that the growth of structure data by itself would lead to even larger deviations in the free parameters; however, the Planck prior is instrumental in restricting their range, notably that of $\p10$ which could lead to unacceptable deviations of $M^2$ at high redshifts. In fact, we have verified this prior to imply the models within the 95\% confidence region have $1.00\lesssim M^2(z_\mathrm{cmb})/\mps \lesssim 1.05$. The constraint on $\so$ induces an even higher precision, almost doubled, on $\omo$ thanks to the Planck prior (see Table \ref{tab:paramprecision}). If this prior were not considered, the precision on $\omo$ suffers a reduction by a factor around four.

It is interesting to emphasise further that the \lcdm fiducial values always lie in a corner of the marginalised contours, if not excluded at two sigma, for all the data sets considered. This is the result of two factors. The first, we have mentioned, the suppression of growth, the second is due to the stability conditions eqs. \eqref{Acond} and \eqref{Bcond} which imply that not all values of $\pij$ are allowed and, in fact, \lcdm corresponds to the very limiting case $A=0$, $B=0$. Therefore, we recover that \lcdm stands indeed at the border of the space of viable EFT of DE models as pointed out in \cite{Piazza:2013pua,Perenon:2015sla,Salvatelli:2016mgy,Perenon:2016blf}. Nonetheless, we observe that the $\chi^2$ at the 95\% confidence limit from the constraints from each data sets does not exclude the Planck \lcdm fiducial model. Having \lcdm at the corner of the 95\% contours or even excluded from the latter, thus does not necessarily imply an exclusion at the level of the full likelihood. Due to the highly non-gaussian behaviour of the likelihood, which is mainly due by the stability conditions, one must be wary of marginalisation effects when considering posterior distributions.

\vv
Looking further into Figure \ref{fig:params}, one can observe that constraints from three measurements of $f$ and  three of $\sig$ yield additional growth suppression with respect to those from $\fs$. This does not come as a surprise since the measurement of $f$ from SDSS and the measurements of $\sig$ by VIPERS exclude the Planck \lcdm fiducial model at more than one sigma. As a result, the \mod model constrained with \fps displays lower values of $\so$, and yields more negative values of $p_{10}$. In parallel, the constraining power of the combination \fps on these two parameters is lower than for the $\fs$ data, since the precision is divided by about a factor two (see Table \ref{tab:paramprecision}).

In general combining $f$ and $\sig$ is likely to break degeneracies that $\fs$ would not and, intuitively, should imply an amelioration of the constraints. Mathematically speaking $\sig (z)$ is an integrated quantity where modified gravity effects induced by the EFT couplings through the effective gravitational coupling $\mu$ are partially washed out, averaged over, during the solving of $\sig$ through the second order differential equation \eqref{eq:evolDplus}. On the other hand, the quantity $f$ is more sensitive to the evolution of the clustering of matter, and must effectively be a better tracer of $\mu$ from the first order differential equation \eqref{eq:evolf}. The relation $f={\rm d}\ln D_+/ {\rm d}\ln a$ is indeed manifest of the integrated nature of $\sig$ with respect to $f$. In fact, from specific tests, we have seen that $f$ measurements drive the constraints on $\sig$, and in parallel we found that SDSS constraints are stronger than those of VIPERS. Understanding $f$ to be the best tracer of $\mu$, one can note that one parameter where the combination \fps does better than $\fs$ is $p_{30}$. This parameter plays a role solely in the fifth force contribution $\muff$. Therefore, one can see how the combination \fps with more measurements could lead to constraining also specific modified gravity phenomena further.

In this respect Table \ref{tab:paramprecision} highlights how well the combination \fspfps performs. With respect to that obtained with $\fs$, the precision on $\so$ is increased by more than 10\% and the $\pij$ by 15\% or more. The best improvement is on $p_{30}$ with a precision increased by more than 30\% . Importantly, we also find that as the result of the combination \fspfps and the Planck prior, the precision obtained on $\p10$ is one order of magnitude larger than the other $\pij$'s.

\renewcommand{\arraystretch}{1.4}
\begin{table}[!]
\begin{center}
\begin{tabular}{|c||c|c|c|c|c|}
\hline 
Data Sets               & $\omo$  & $\so$   & $p_{10}$ & $p_{20}$ & $p_{30}$  \\ 
\hline\hline  
$\fs$                   & $38.72$ & $24.16$ & $32.87$  & $1.20$   & $1.95$    \\
\hline 
$f+\sigma_8$            & $39.24$ & $10.89$ & $17.04$  & $1.31$   & $2.18$ 	\\	
\hline 
$\fs+f+\sig$            & $41.05$ & $26.88$ & $37.58$  & $1.49$   & $2.59$    \\	
\hline 
$f+\sig \; / \; \fs$    & $1.01$  & $0.45$  & $0.52$   & $1.09$   & $1.12$    \\	
\hline 
$\fs+f+\sig \; / \; \fs$ & $1.06$  & $1.11$  & $1.15$   & $1.24$   & $1.33$    \\
\hline
\end{tabular} 
\caption{The precision on the constraints, $i.e.$ the inverse of the 68\% marginalised confidence interval length, from each data set. The last two lines of the table correspond to the ratio of the precisions as indicated by the label in the first column.}
   \label{tab:paramprecision}
\end{center}
\end{table}

\vv
We have not yet discussed the parameter $p_{20}$ and it deserves particular attention. It characterises the one coupling, $\m22$, which does not take part in the computation of the observables when the QSA is considered. However, $\m22$ does play a significant role in the no ghost condition eqs.\eqref{Acond} and regulates thus the size of the space of viable models and the propagation speed of scalar modes \cite{Piazza:2013pua}. It is a priori unexpected therefore to be able to constrain such a parameter through a cosmological analysis when the QSA is considered. Note that the coupling $\m22$ is closely related to the kinetic coupling of the ``$\alpha$-basis", an equivalent description of Horndeski theories with the EFT of DE at the perturbation level \cite{Bellini:2014fua},
\begin{equation}\label{eq:alphak}
\alpha_K = \frac{2\mathcal{C}+4\m22}{H^2}\; .
\end{equation}
In the literature, the kineticity has hence been generally set to a given value \cite{Bellini:2015xja,Gleyzes:2015rua,Salvatelli:2016mgy,Alonso:2016suf,Mancini:2018qtb,Reischke:2018ooh} so as to avoid losing predictability. The constraints considering all the relevant cosmological probes obtained on the 1D parametrisation of the $\alpha$ couplings are virtually unaffected by the tested values of $\alpha_K$ (see Figure 3 in \cite{Bellini:2015xja}). The difficulty of constraining it was also recently shown to arise from its observable contribution being below cosmic variance \cite{Frusciante:2018jzw}. However, interestingly, the best fit obtained in \cite{Bellini:2015xja} for each value of $\alpha_K$ does induce significantly different predictions of the growth function $\fs$ at low redshifts depending on the combination of data sets (see Figure 2 in \cite{Bellini:2015xja}). 

\vv
The best fit itself is not statistically significant enough to draw strong conclusions. It is however interesting to note that only once RSD data is considered on top of CMB and other cosmological data sets, do the best-fit predictions in \cite{Bellini:2015xja} on $\fs$ for the different values of $\alpha_K$ match. This might be a hint of accrued sensitivity of the combination of the growth of structure data and viability requirements on the kineticity with respect to other cosmological probes. As a matter of fact, we do not obtain a flat posterior for $p_{20}$ as should have been expected if the constraining power were too low. One must understand this constraint to be indirect. The parameter $p_{20}$ does not modify the likelihood values but only impacts the viability conditions $A>0$ and $c_s^2>0.1$ imposed as hard priors in our analysis. It regulates the space of viable models. In our results, a skewed posterior towards \lcdm is highlighted where negative values of $p_{20}$ are cut out by the conditions $A>0$, while larger values are disfavoured by the prior on $c_s^2$ which implies $B>0.1\,A$. In summary, the constraint on $p_{20}$ can be schematically understood as follows. The other $\pij$'s being constrained by the data, only a small window is left for $p_{20}$ to vary within the stability conditions. We thus conclude that it is possible to obtain, although indirectly, constraints on $\m22$ when the QSA is considered thanks to the back reaction of the viability conditions. 

\subsection{LSS functions and parametrisation dependency}\label{sec:obs1}

\begin{figure}[!]
\begin{center}
1D \hskip80mm 2D
\vskip1mm
\includegraphics[scale=0.52]{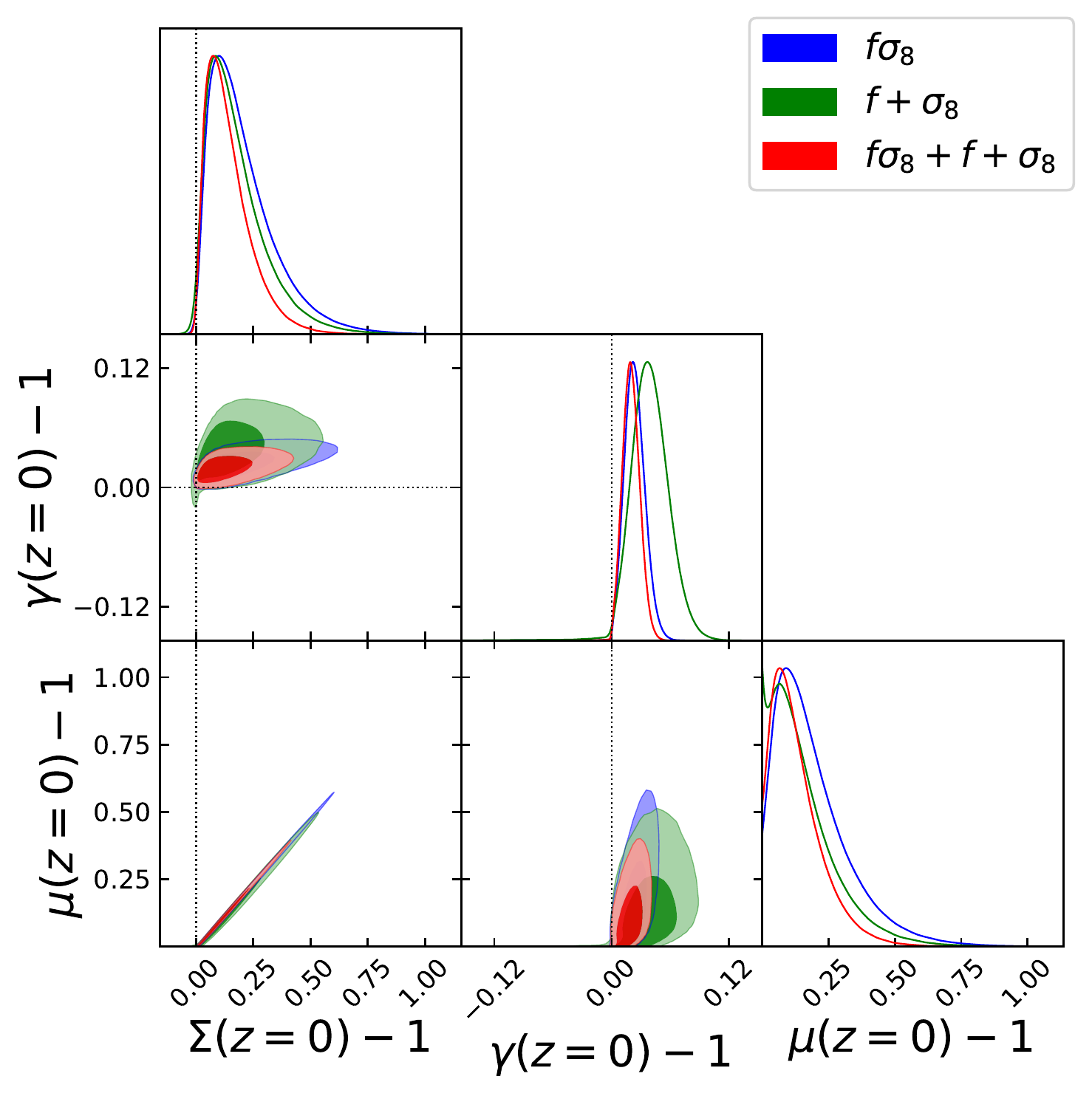}
\hskip4mm
\includegraphics[scale=0.52]{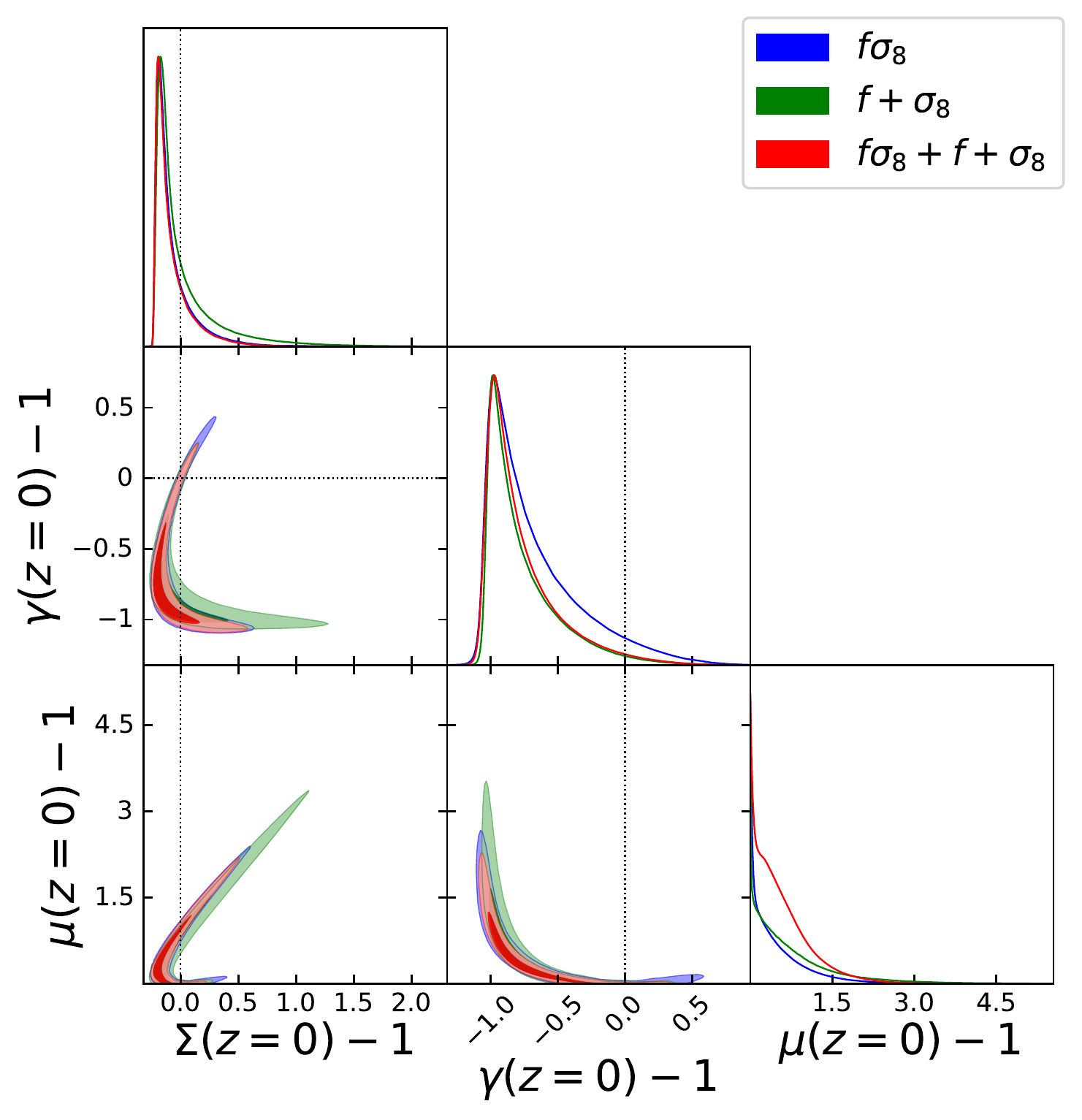}
   \caption{The constraints obtained on the LSS observables $\mu$, $\Sigma$ and $\gsp$ from the data sets are displayed for the 1D (\emph{left panel}) and 2D (\emph{right panel}) parametrisations. In each plot, the standard model is represented by dotted lines.}
   \label{fig:currentobs}
\end{center}
\end{figure}

The LSS observables $\mu$, $\gsp$ and $\Sigma$ are key in detecting deviations from the standard model \cite{2010PhRvD..81j4023P,Song:2010fg,Simpson:2012ra,Ade:2015rim,Ferte:2017bpf,Aghanim:2018eyx} and the correlations of these observables have been thoroughly studied in the context of Horndeski theories \cite{Perenon:2015sla,Salvatelli:2016mgy,Perenon:2016blf,Peirone:2017ywi}. 

Assessing how the previous constraints on the parameters translate onto these observables, one striking feature displayed in Figure \ref{fig:currentobs} (left panel) is that the favoured values of $\gsp$ are one order of magnitude smaller than $\mu$ or $\Sigma$. From this severe constraint of $\gsp$ around unity, with values slightly larger than one favoured, $\mu$ and $\Sigma$ are hence correlated. An additional feature one can observe is that for $\mu$ or $\Sigma$ the \fps combination yields better constraints than with $\fs$. The precision obtained on $\mu$, $\Sigma$ is increased by respectively 20\% and 10\% whereas $\gsp$ suffers a 45\% percent reduction. As a result, we find the combination \fspfps to increase the precision on $\mu$, $\Sigma$ by more than 40\% and more than 15\% for $\gsp$ with respect to the $\fs$ data set. Ref. \cite{Salvatelli:2016mgy} constrained the 1D parametrisation with CMB data. In their analysis $c_T$ is not fixed to one while $\m22$ is set to zero, which renders the comparison with our results somewhat biased. Nevertheless, it is clear that growth of structure has a significantly superior constraining power for the free parameters and the LSS observables than CMB data. This makes the constraints we discuss in this present paper more conservative.  

\vv
We must emphasise also that the constraints are model-dependent yet this model dependency is enlightening to discuss. The tightness of constraints we have discussed up to now must be put in perspective with the amount of freedom given to a model. While limiting the amount of parameters serves the purpose of clearly highlighting the constraining power of each data set, it by no means necessarily implies a faithful description of the observable predictions of the model. Once we allow more freedom in the coupling functions by considering the 2D parametrisation, the constraints on the LSS observables change significantly as displayed in Figure \ref{fig:currentobs} (right panel). In short, one can see $\gsp$ to no longer be bounded close to unity and values smaller than one are now favoured. The constraints on the three observables show the U-shape previously found in \cite{Perenon:2016blf}. This is due to the fact that $\mu$ and $\gsp$ share a common term: the denominator of the former is a factor in the numerator of the latter (see \cite{Perenon:2016blf} for more details).

Looking further into the constraints from each data set, the picture is now opposed: \fps constrains $\gsp$ better than $\mu$. Consequently, we find the \fspfps combination with respect to $\fs$ to improve the precision respectively on $\gsp$ by 35\% and $\Sigma$ by 5\%, whereas the precision on $\mu$ suffers a 15\% reduction. These changes originate from the fact that the favoured values of the two parameters of the same coupling can be large but compensate each other and allow for similar variations of the couplings. This turns out to be critical for the evolution of the bare Planck mass for instance. In practice, with the 2D parametrisation we obtain the mean value $\p10 \approx -6.152_{-4.096}^{3.777}$ (95\% c.l., see Table \ref{tab:paramlimits_gdotg}) for \fspfps as opposed to $p_{10}= -0.031_{-0.025}^{+0.026}$ (95\% c.l.) in the 1D case. In the latter case, $\p10$ was thoroughly constrained by the Planck prior, whereas in the former both $p_{1j}$ can absorb this prior. We emphasise that the values of $\so$ barely change between the parametrisations. 

\vv
We therefore see that the evolution of the bare Planck mass is instrumental to the effectiveness of the growth of structure data and more precisely for the \fps splitting. When large variations are allowed, the contribution of $\musc$ to $\mu$ can cancel the contribution of $\muff$, in other words, an increase induced by the latter is likely to be counter-balanced by a suppression induced by the former. As soon as the variation of $M^2$, hence $\musc$, is limited, as described by the 1D parametrisation, growth of structure data and notably $f$ is able to constrain $\mu$ more stringently hence $\muff$ in particular. In parallel, an important difference to keep in mind is that the gravitational slip parameter on the contrary does not contain the $\mps/M^2$ overall factor and the analysis is thereby reversed: the \fps combination is less powerful than $\fs$ when $M^2$ has small variations.    

\section{Reducing the space of viable models}

An important point from the previous section is the caveat on the generality of a parametrisation. When investigating statistical inferences, one has to find the meaningful middle ground between small parameter spaces for effective constraints and faithful modelling. The use of simple parametrisations, as we have done here with the 1D choice, allows us to thoroughly assess and compare the constraining power of various cosmological probes on the parameter space \cite{Bellini:2015xja,Salvatelli:2016mgy}. However when turning to observables predictions, additional care must be taken so as to ensure that the predictions are representative and unbiased. In the context of Horndeski theories this requires more involved parametrisations \cite{Perenon:2015sla,Linder:2015rcz,Perenon:2016blf,Linder:2016wqw,Gleyzes:2017kpi,Espejo:2018hxa} and can lead effectively to disappointingly large constraints. Nevertheless, one must utilise physical priors to complement statistical inferences and overcome the large freedom needed for a correct description of a model. While stability conditions stand as theoretical priors, one should also consider observational ones. The gravitational theory provided by the EFT of DE allows one to make predictions on a large range of scales. As discussed in Section \ref{sec:characLSS}, the effective gravitational coupling $\mu$ is a quantity characterising modifications of gravity on linear cosmological scales and impacts directly the growth of structure. This however does not mean that all the modifications of gravity acting in $\mu$ encode effects only on such scales. As a matter of fact, we have seen this coupling to be the result of two contributions: the fifth force induced by the extra scalar field $\muff$ and an intrinsic modification of gravity that is un-screenable, $\musc$ (see eq. \eqref{eq:mu}). Therefore, in this case certain astrophysical constraints can be naturally applied to r $\musc$ and equivalently the bare Planck mass $M^2$. Note that the use of the Hulse-Taylor pulsar led in this way to a bound on the speed of tensor modes to unity at the $10^{-2}$ level in \cite{Jimenez:2015bwa}.

The variation of the bare Planck mass is a critical point when considering growth of structure constraints and its variation is severely bounded observationally. To apply the bounds on the variation of Newton's constant onto the variation of the bare Planck mass at present time, one must make use of the discussion around eq. \eqref{eq:musc}. Using the definitions $G_\mathrm{sc}(z=0) = 1/(8\pi M(z=0)^2)\equiv \gn$ and $\mu_1(t)={\rm d}\ln M^2(t)/{\rm d}\ln t$, one can derive the relation
\begin{equation}\label{eq:vargn}
\frac{\dot{\gn}}{\gn}=\frac{\dot{G_\mathrm{sc}}}{G_\mathrm{sc}} \bigg|_{z = 0}= \frac{{\rm d}\musc}{{\rm d} t}\bigg|_{z = 0}= -\mu_1(z=0)\;.
\end{equation}
There are numerous bounds on the variation of Newton's constant \cite{Uzan2011} and the strongest constraint has been obtained from Mars ranging data \cite{2011Icar..211..401K}, which gives a conservative bound  $|\dot{G}_\mathrm{N} / G_\mathrm{N}|\lesssim 1.6 \times 10^{-13}\;{\rm year}^{-1}$. Remaining conservative also in terms of measurements of $H_0$, we translate this bound into $ |\dot{G}_{\rm N}/\gn|< 0.002\;H_0$. 
Imposing this conditions on our models through \eqref{eq:vargn} induces from the parametrisation eq. \eqref{parammu1} the stringent bound on the parameter $p_{10}$:
\begin{equation} \label{eq:boundgdot}
\abs{\mu_1(z=0)} \,= \, \abs{p_{10}} < 0.002\ .
\end{equation}
This $\gdot$ prior forces thus the free parameter $p_{10}$ to be extremely close to its corresponding \lcdm value $p_{10}=0$.

\subsection{Enhancing the power of data with $\gdot$}

\begin{figure}[!]
\begin{center}
\includegraphics[scale=0.4]{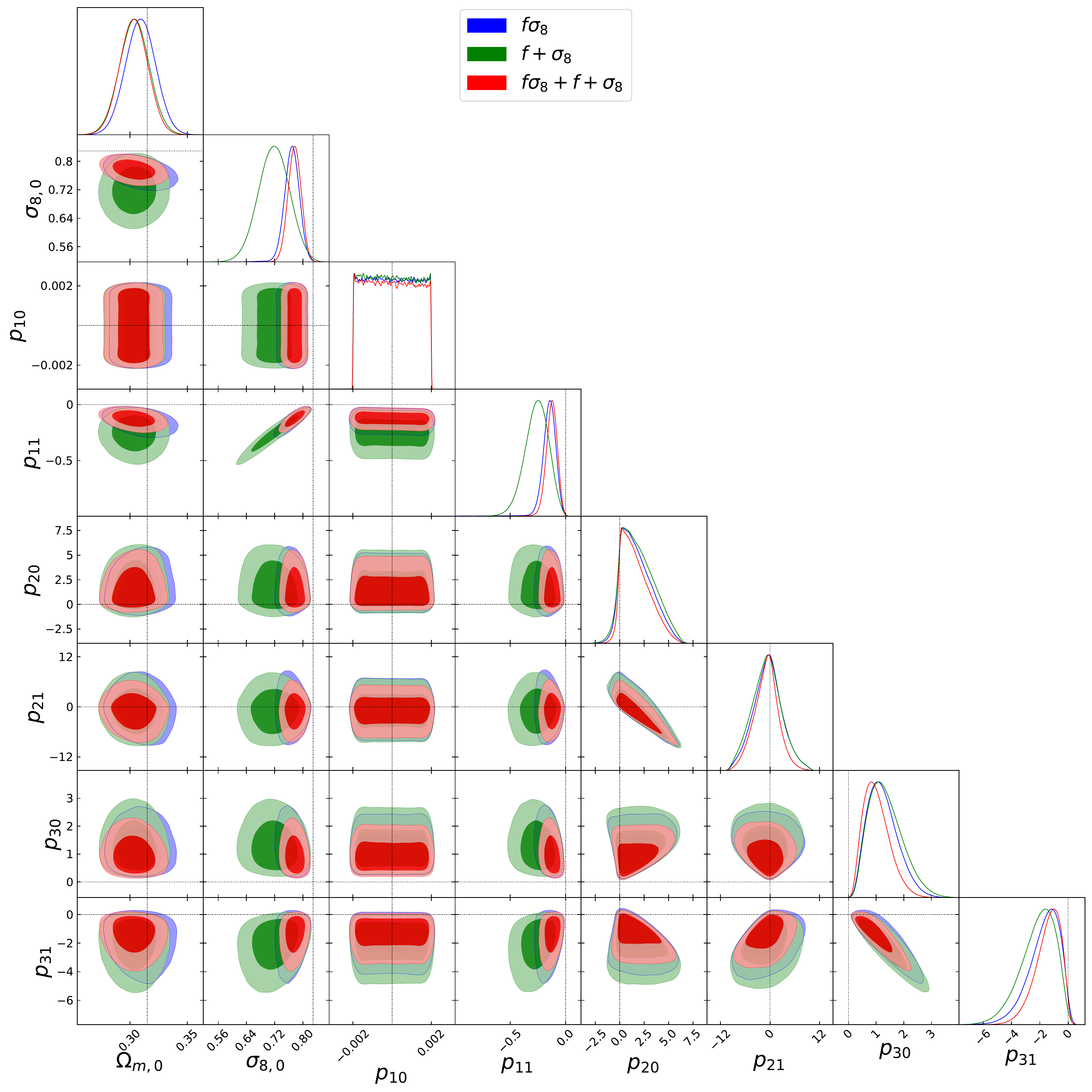}
   \caption{The constraints obtained on the free parameters of the \mod model in the 2D parametrisation. Compared to the results of Figure \ref{fig:params}, here, the $\gdot$ prior has been considered in all the likelihoods.}
   \label{fig:params_gdot}
\end{center}
\end{figure}

\renewcommand{\arraystretch}{1.4}
\begin{table}[!]
\begin{center}
\begin{tabular}{|c||c|c|c|c|c|c|c|c|}
\hline 
Precision                             & $\omo$  & $\so$   & $p_{10}$  & $p_{11}$ & $p_{20}$ & $p_{21}$ & $p_{30}$ & $p_{31}$  \\ 
\hline\hline  
$\fs+f+\sig$                          & $38.41$ & $21.92$ & $0.25$    & $0.07$   & $0.05$   & $0.02$   & $0.25$   & $0.07$  \\
\hline 
$\fs+f+\sig+\gdot$                    & $40.96$ & $24.22$ & $367.64$  & $10.35$  & $0.38$   & $0.18$   & $1.11$   & $0.59$  \\	
\hline 
$\fs+f+\sig+\gdot \; / \; \fs+f+\sig$ & $1.07$  & $1.24 $ & $1461.67$ & $149.38$ & $7.95$   & $7.66$   & $4.37$   & $8.08$  \\
\hline
\end{tabular} 
\caption{The precision on the constraints, $i.e.$ the inverse of the 68\% marginalised confidence interval length from the \fspfps data set with and without the $\gdot$ prior. The last line of the table corresponds to the ratio of the precisions of the two cases.}
   \label{tab:paramprecision_gdotg}
\end{center}
\end{table}

\vv
Through this stringent cut, the growth of structure data no longer has enough power to constrain $\p10$ within this window and the posterior is flat; however, this cut propagates to the other $\pij$'s rendering a significant improvement in their constraints (see Figure \ref{fig:params_gdot}). Considering the full combination $\fs+f+\sig$, the precision on the $\pij$'s is at least quadrupled when the $\gdot$ prior is added (see Table \ref{tab:paramprecision_gdotg}). The case of $\lbrace p_{10},\,p_{11} \rbrace$ is striking, these two being the parameters of $\mu_1$, hence $\musc$, it is not surprising to have obtained the best improvement on $\p10$ and $p_{11}$, once the $\gdot$ prior is added.

With the bare Planck mass bounded from this observational prior, we recover what was discussed previously. The ability of the combination \fps to trace the evolution of $\mu$ is enhanced significantly and as a result this combination does as well as the $\fs$ data set, except for $p_{11}$ and $\so$ (see Figure \ref{fig:params_gdot}). The latter is unsurprising since the that relative error of the SDSS $\sig$ measurement is quite large compared to that of the measurements of $\fs$ at low redshifts.

\vv
Furthermore, $f$ and in parallel the SDSS measurement were found to drive the constraints in the \fps combination in the previous section. However, it is difficult to judge how this relates to the precision of the data. The SDSS data points are a release at low redshifts where $f$ is more precise than $\sig$ while it is the opposite for VIPERS; a much higher redshift release with $\sig$ more precise than $f$. To assess this uncertainty, we apply a mock test where we keep the $\gdot$ prior, the redshift and central values of the SDSS and VIPERS measurements, but we enforced a relative error of 10\% on all the measurements. We find again that the low redshift measurements have more impact and in parallel $f$ is more sensitive to the $\pij$'s than $\sig$. This confirms that $f$ is a better tracer of the evolution of $\mu$ as understood intuitively and widely debated previously. 

Nevertheless, it is interesting to point out that the addition of the $\sig$ data improves the constraints on $\lbrace \so,\,p_{11} \rbrace$ while virtually not affecting those on $\lbrace p_{30},\,p_{31} \rbrace$. This might be a hint that the constraining power of $\sig$ is mostly on the $\musc$ component, the one responsible for suppressed growth. Not surprisingly in fact, since the amplitude of the $\sig(z)$ predictions are mostly driven by $M^2$. The relative precision needed for the addition of the $\sig$ data to produce at least the same precision on the parameters as the addition of $f$ data with 10\% relative error, must be increased around a factor around 10. An additional mock test shows that the relative error on the three measurements of $\sig$ must be around 1\%. 

\begin{figure}[!]
\begin{center}
\includegraphics[clip, trim = 0cm 27cm 0cm 0cm, scale=0.4]{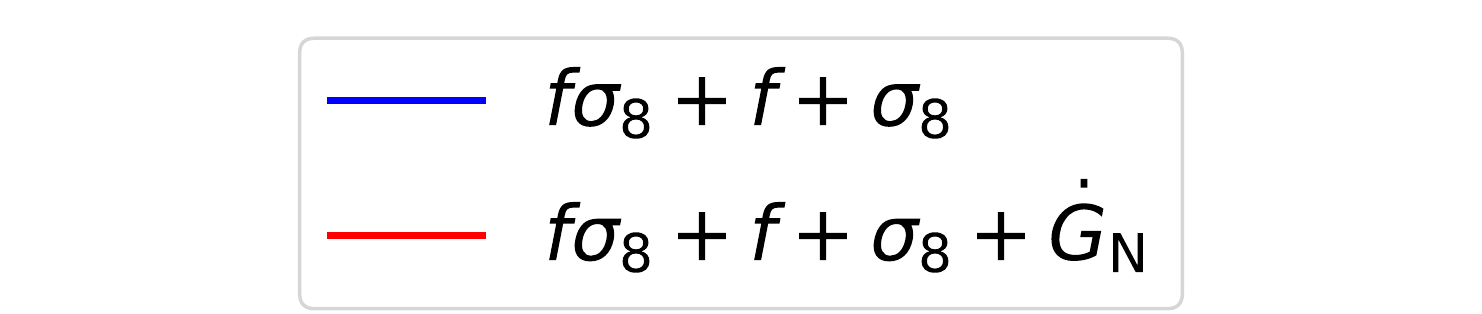}
\vskip4mm
\includegraphics[clip, trim = 0cm 2cm 0cm 0cm, scale=0.2]{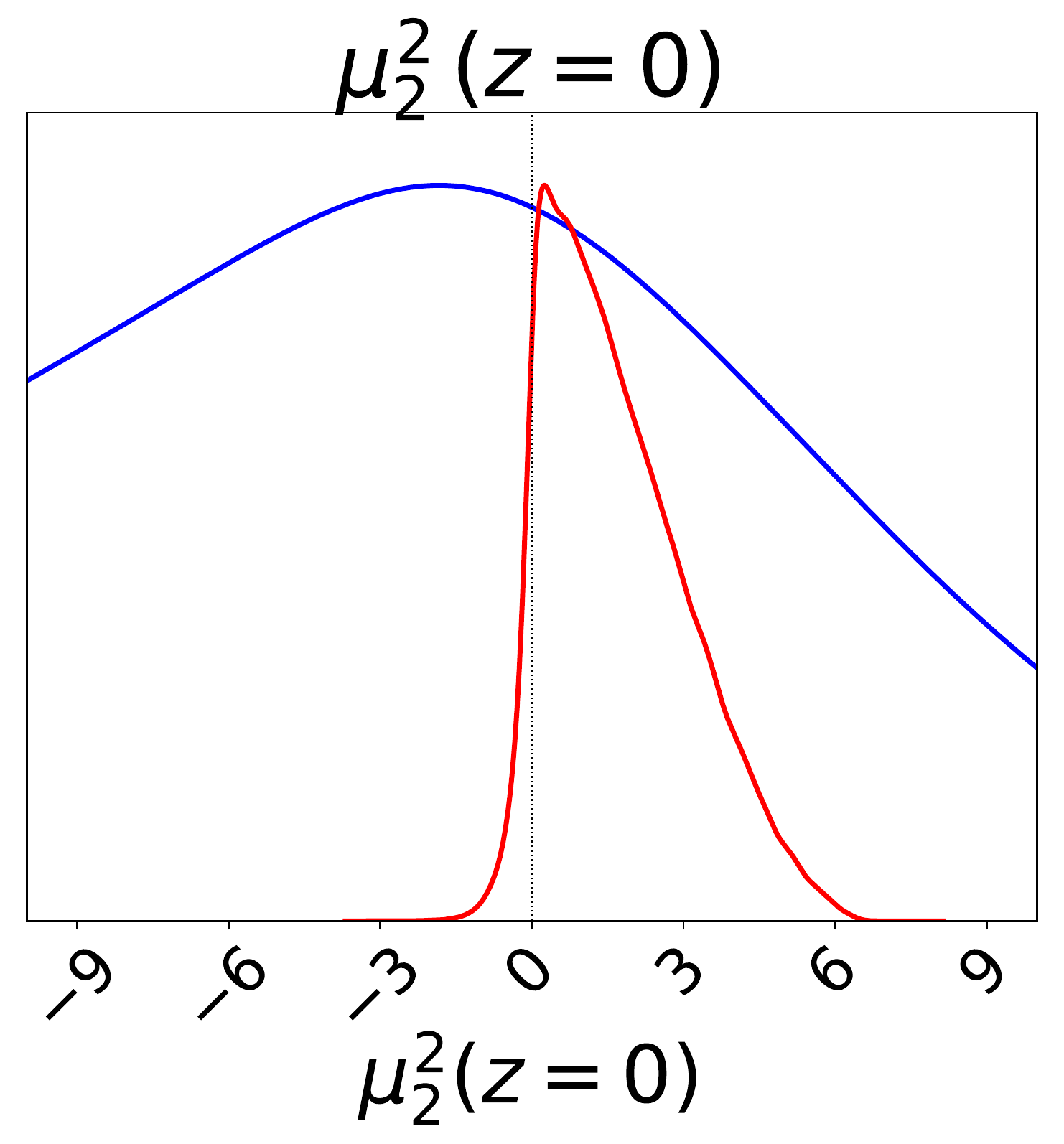}
\hskip3mm
\includegraphics[clip, trim = 0cm 2cm 0cm 0cm,scale=0.2]{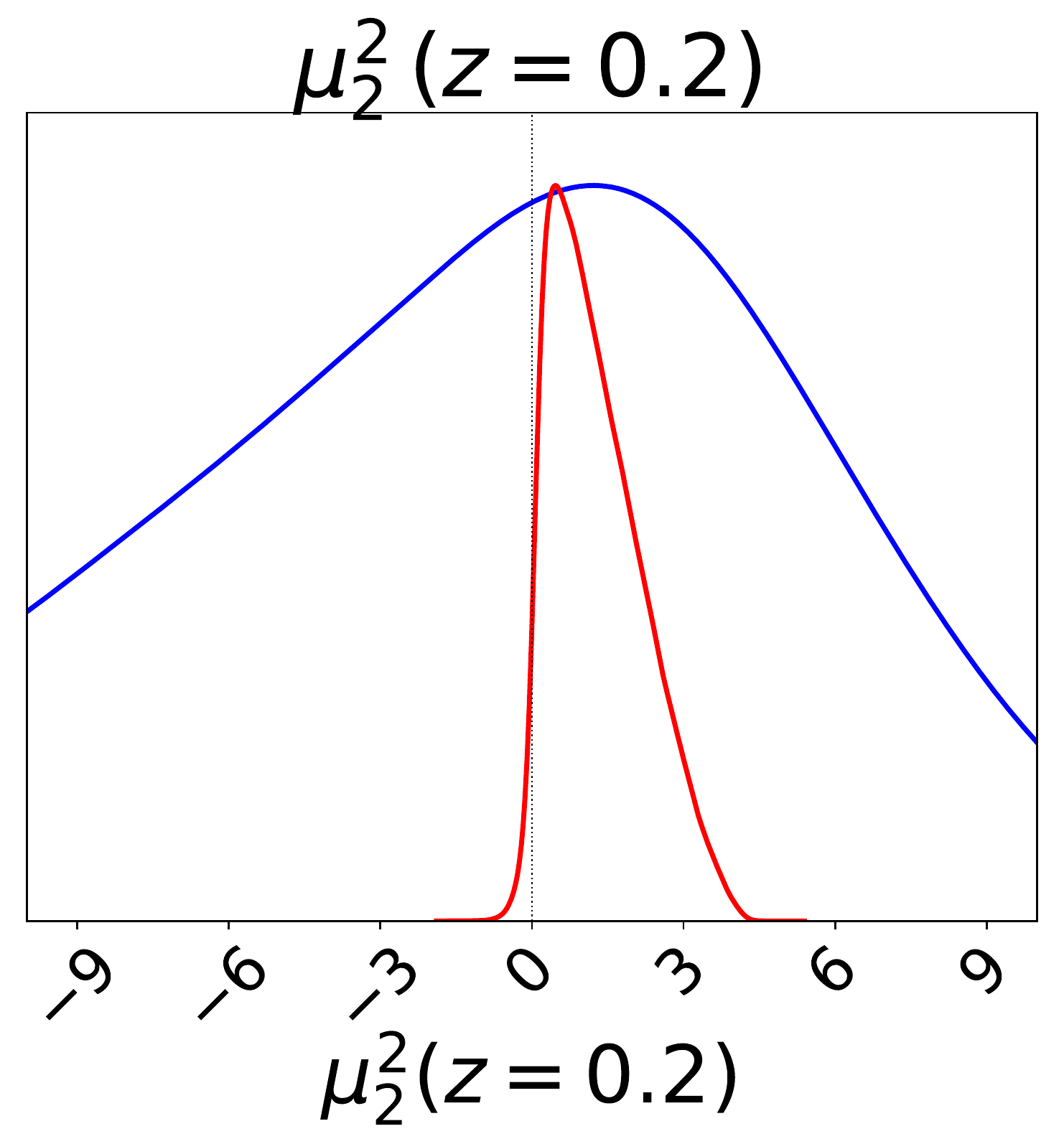}
\hskip3mm
\includegraphics[clip, trim = 0cm 2cm 0cm 0cm,scale=0.2]{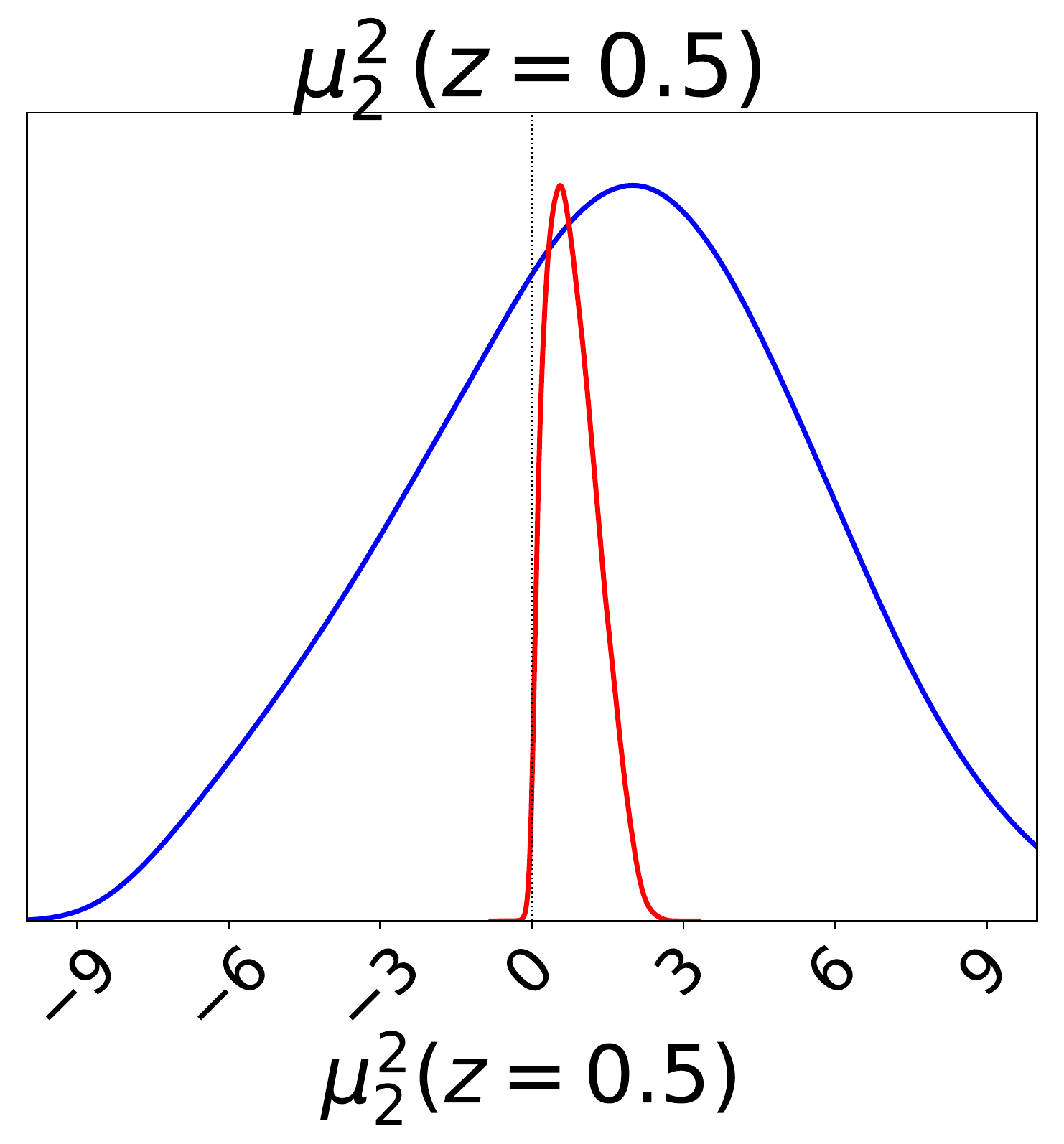}
\hskip3mm
\includegraphics[clip, trim = 0cm 2cm 0cm 0cm,scale=0.2]{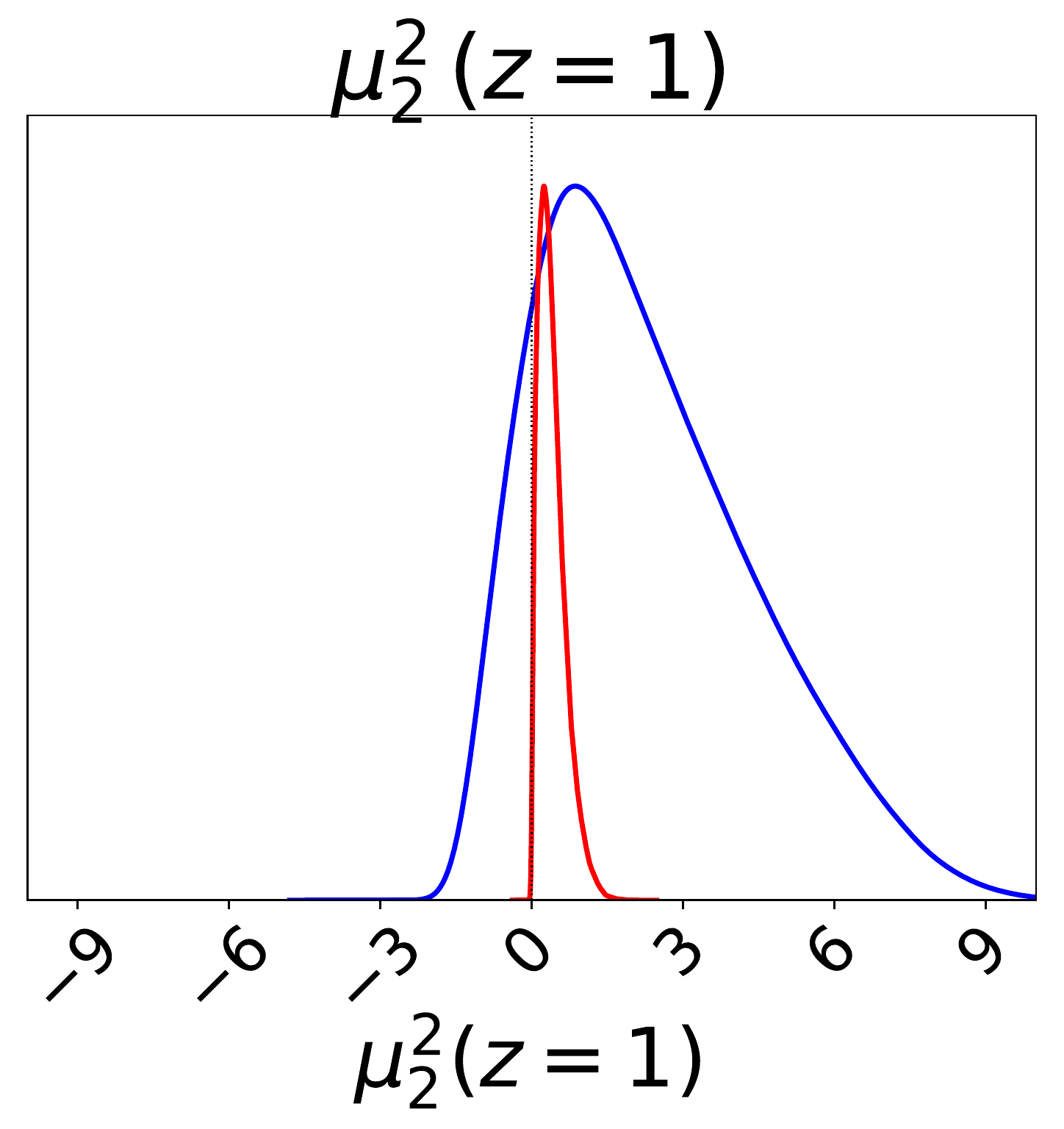}
\vskip3mm
\includegraphics[clip, trim = 0cm 1.9cm 0cm 0cm,scale=0.2]{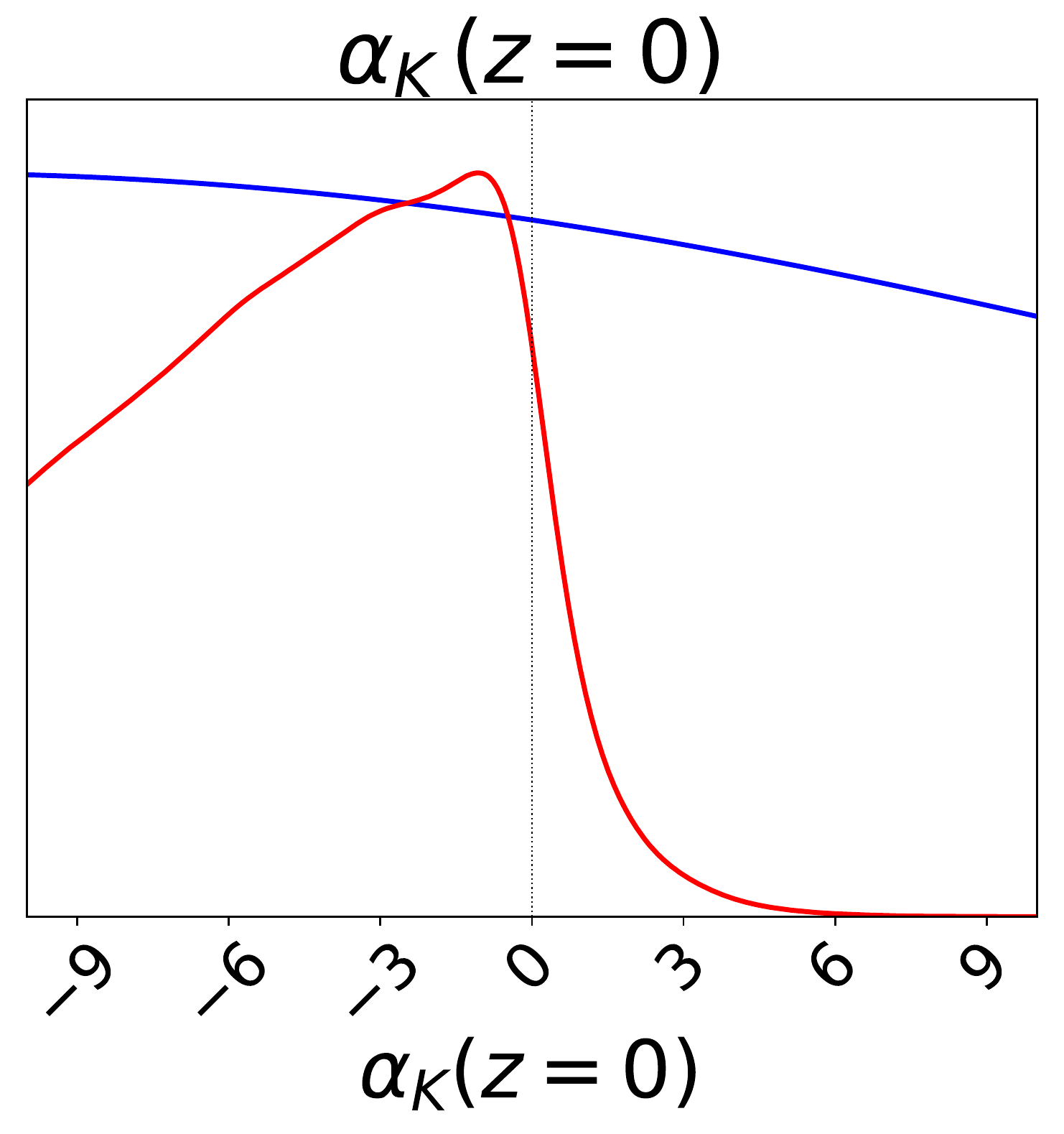}
\hskip3mm
\includegraphics[clip, trim = 0cm 1.9cm 0cm 0cm,scale=0.2]{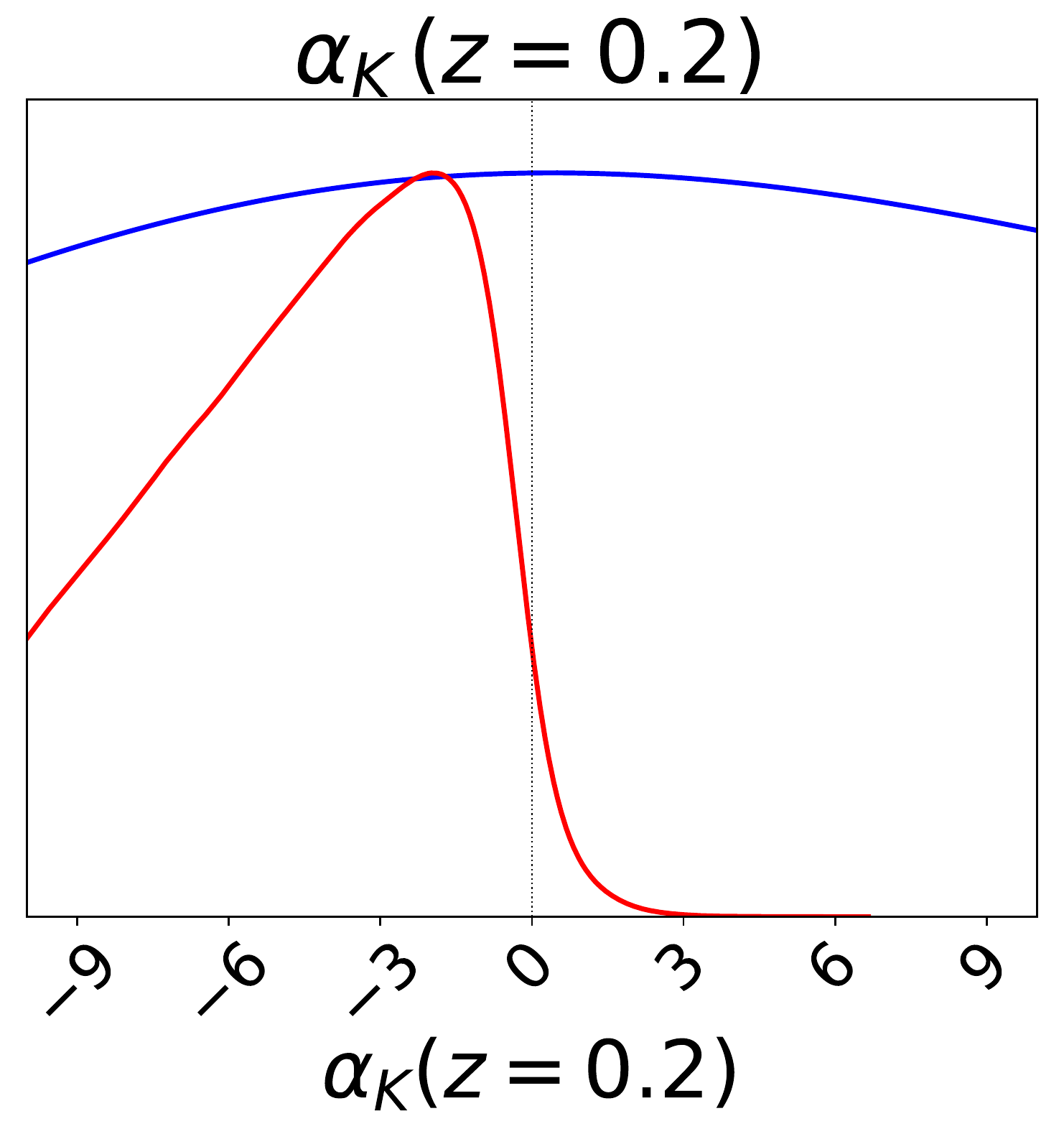}
\hskip3mm
\includegraphics[clip, trim = 0cm 1.9cm 0cm 0cm,scale=0.2]{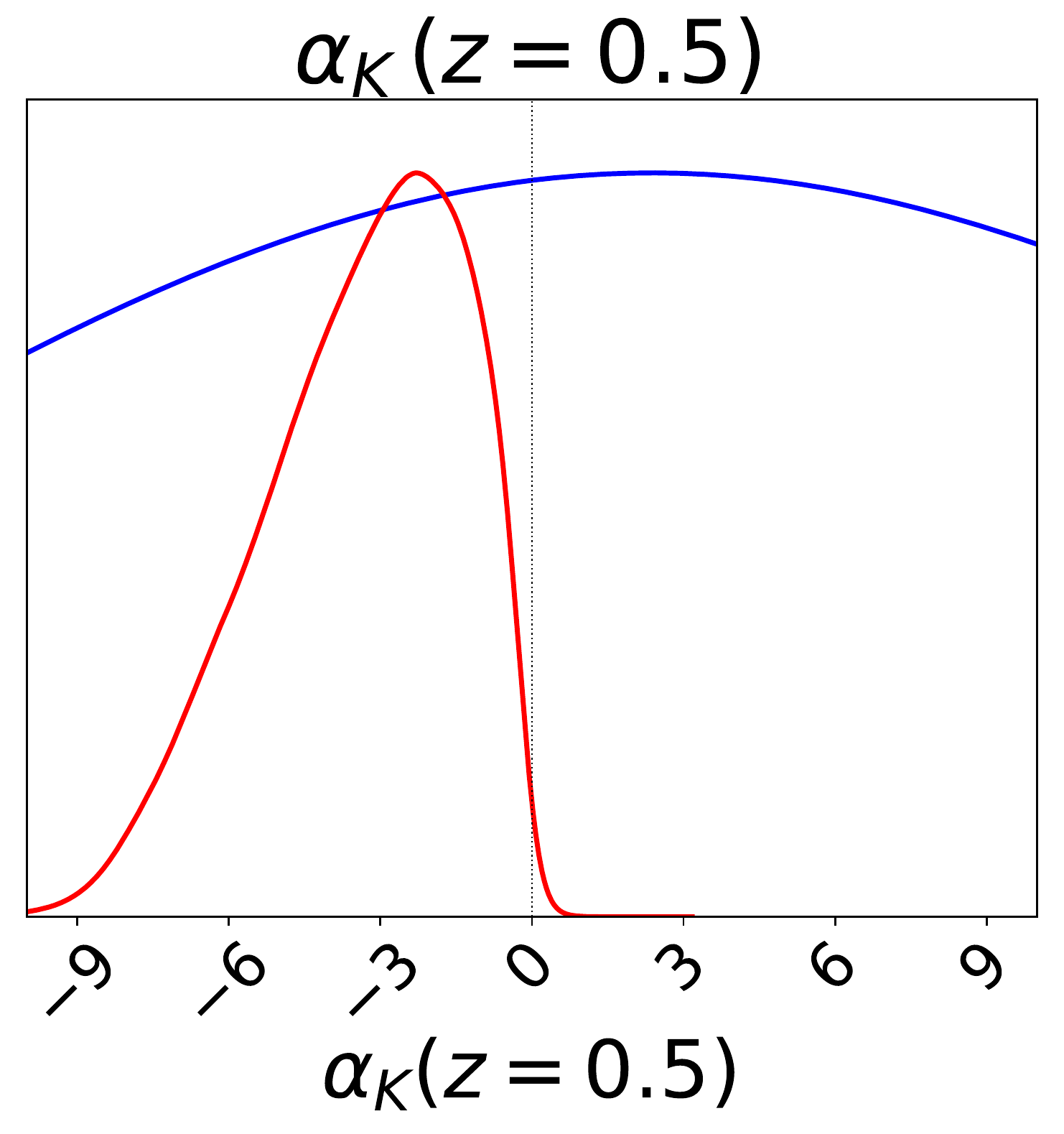}
\hskip3mm
\includegraphics[clip, trim = 0cm 1.9cm 0cm 0cm,scale=0.2]{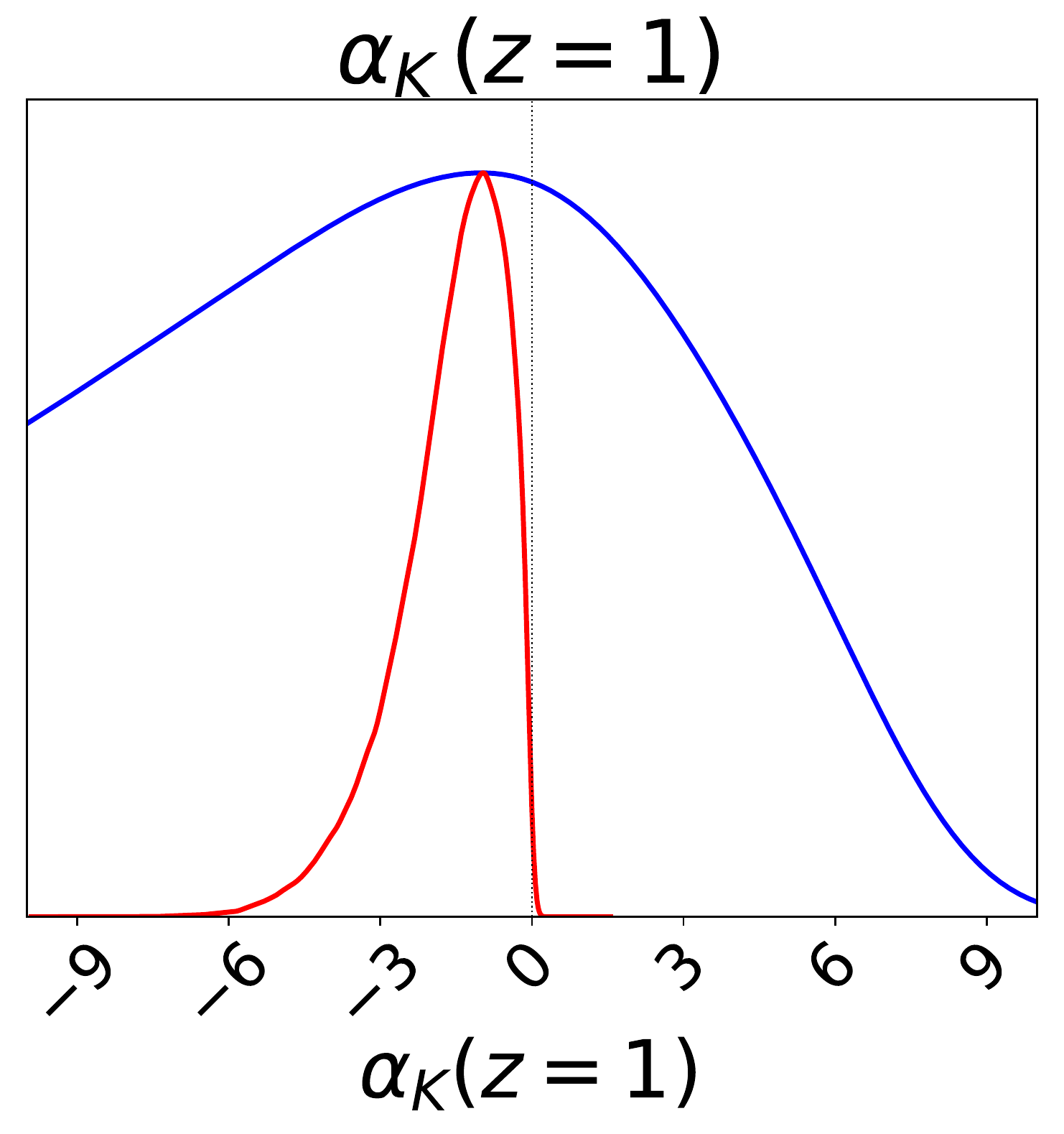}
   \caption{Redshift evolution of the 1D posterior normalised to peak at one of $\m22$ and $\ak$ from considering the \fspfps combination with and without the $\gdot$ prior.}
   \label{fig:alphak}
\end{center}
\end{figure}

\vv
Studying the constraints on the coupling $\m22$, we emphasise that without the $\gdot$ prior $\lbrace p_{20},\, p_{21} \rbrace $ would effectively be unconstrained (see Table \ref{tab:paramlimits_gdotg}) showing that this prior is instrumental to constrain possible variations of $\m22$ when more freedom is allowed through the 2D parametrisation. This highlights how the $\gdot$ condition acts as a stringent cut on the space of viable models and completes efficiently the set of theoretical priors given by the viability conditions. We observe in Figure \ref{fig:alphak} the constraints tighten with redshift due to the $1-\om$ switching off and the Planck prior, where the $\gdot$ prior is crucial to constrain $\m22$ within an acceptable range across redshifts. On the other hand, the kineticity coupling $\ak$ displays much larger posteriors due to the coefficients in eq. \eqref{eq:alphak} and the contribution of $\mathcal{C}$. The $\gdot$ prior is only able to constrain its amplitude within the $\left[-10, 10\right]$ window at $z \gtrsim 0.5 $. The width of the constraints on $\alpha$ is consistent with the fact that fixing $\ak$ to a given value has little effect on cosmological constraints in previous studies.     

\vv
To conclude on the parameter constraints, we emphasise that it is indeed not novel that RSD data favours models predicting lower growth compared to the standard model at low redshifts. We see this clearly from the constraint on $\so$ whatever the likelihood setup. However, an important new finding here is that this remains true despite the application of the $\gdot$ prior. This could be surprising, since the $\gdot$ prior acts as a strong constraint on the derivative of $M^2(z=0)$, or equivalently $\musc(z=0)$, the sole contribution to $\mu$ producing weaker gravity. Not only does the suppression of growth remain, but the mean value of $\so$ is lowered further by the application of this prior (see Table \ref{tab:paramlimits_gdotg}). Essentially, the models have a period in which $\musc$ is slightly lower than unity, even if very mildly, spanning a sufficiently large redshift period to produce lower growth than \lcdm.

\subsection{Redshift evolution of the LSS functions}

\begin{figure}[!]
\begin{center}
\includegraphics[clip, trim = 0cm 11cm 0cm 0cm,scale=0.34]{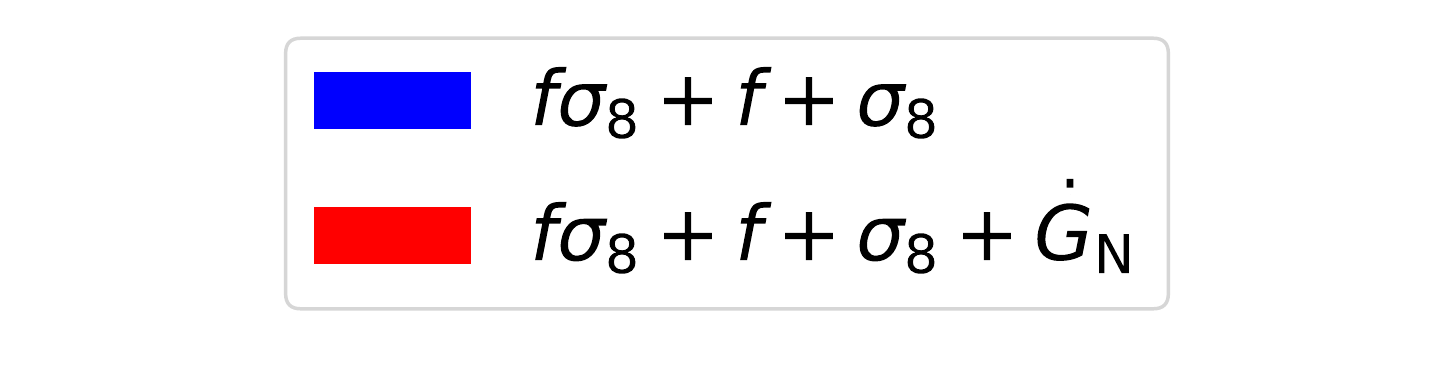}
\vskip2mm
\includegraphics[clip, trim = 0.05cm 0.05cm 0.05cm 0.05cm,scale=0.17]{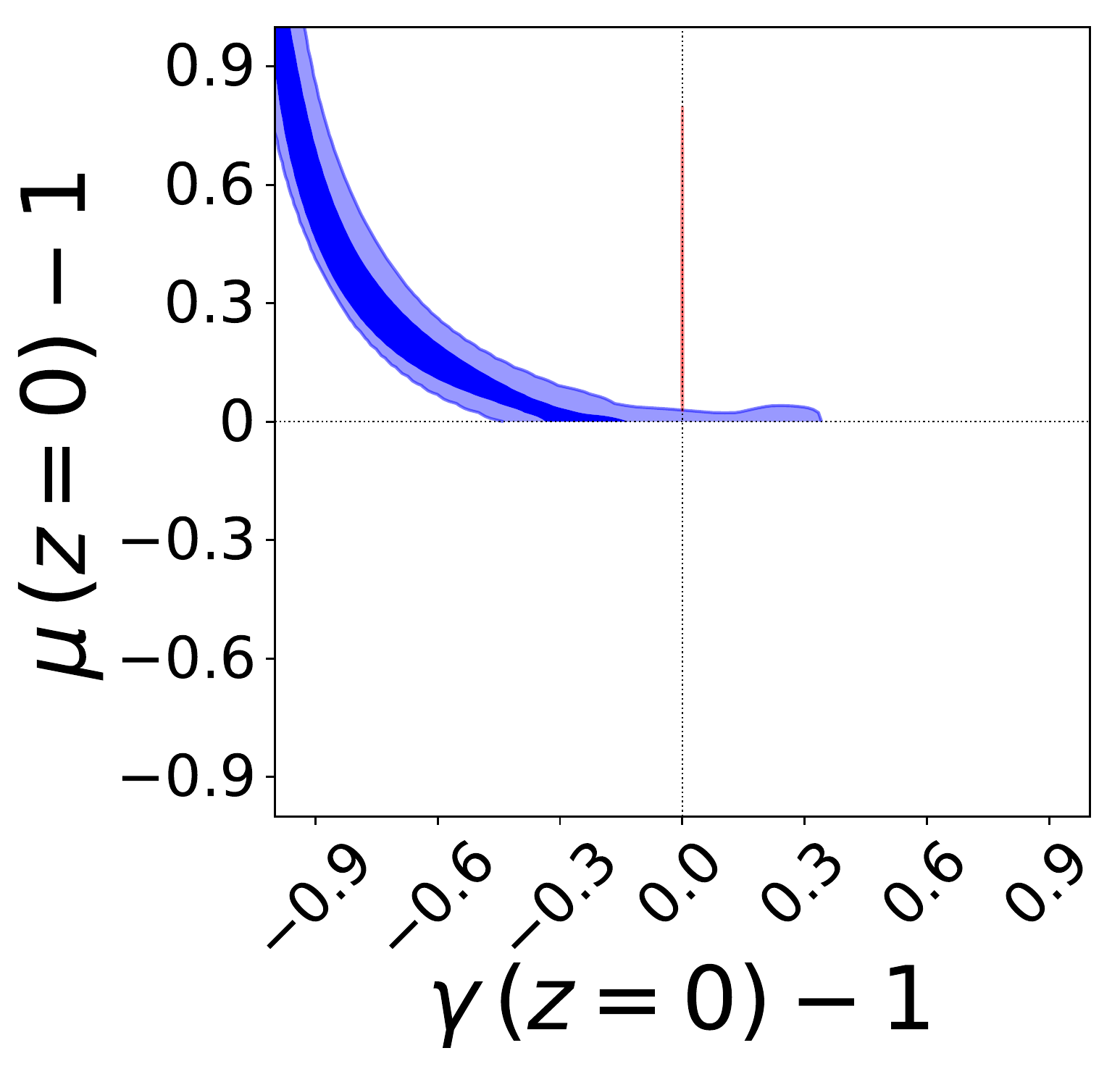}
\includegraphics[clip, trim = 0.05cm 0.05cm 0.05cm 0.05cm,scale=0.17]{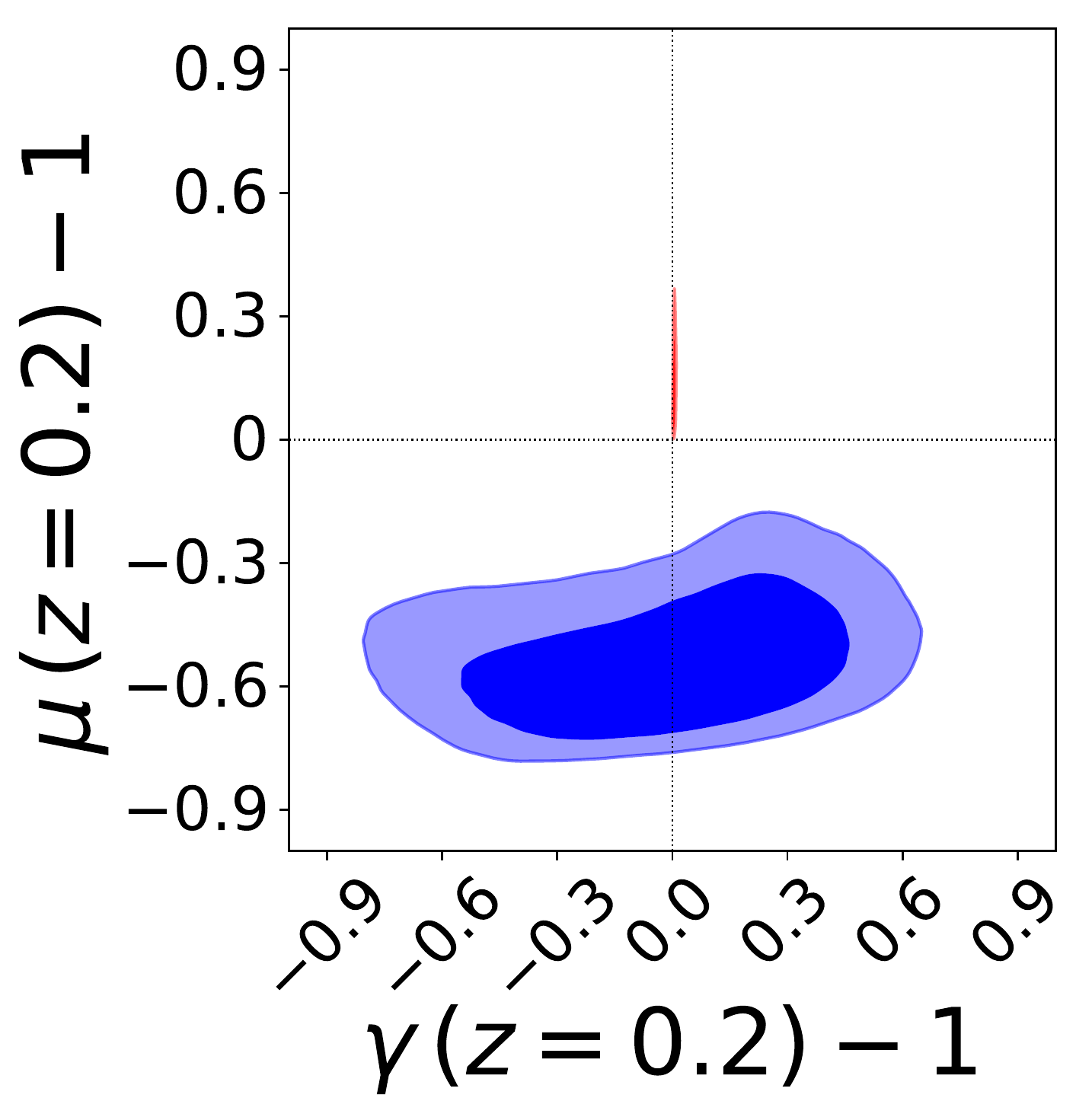}
\includegraphics[clip, trim = 0.05cm 0.05cm 0.05cm 0.05cm,scale=0.17]{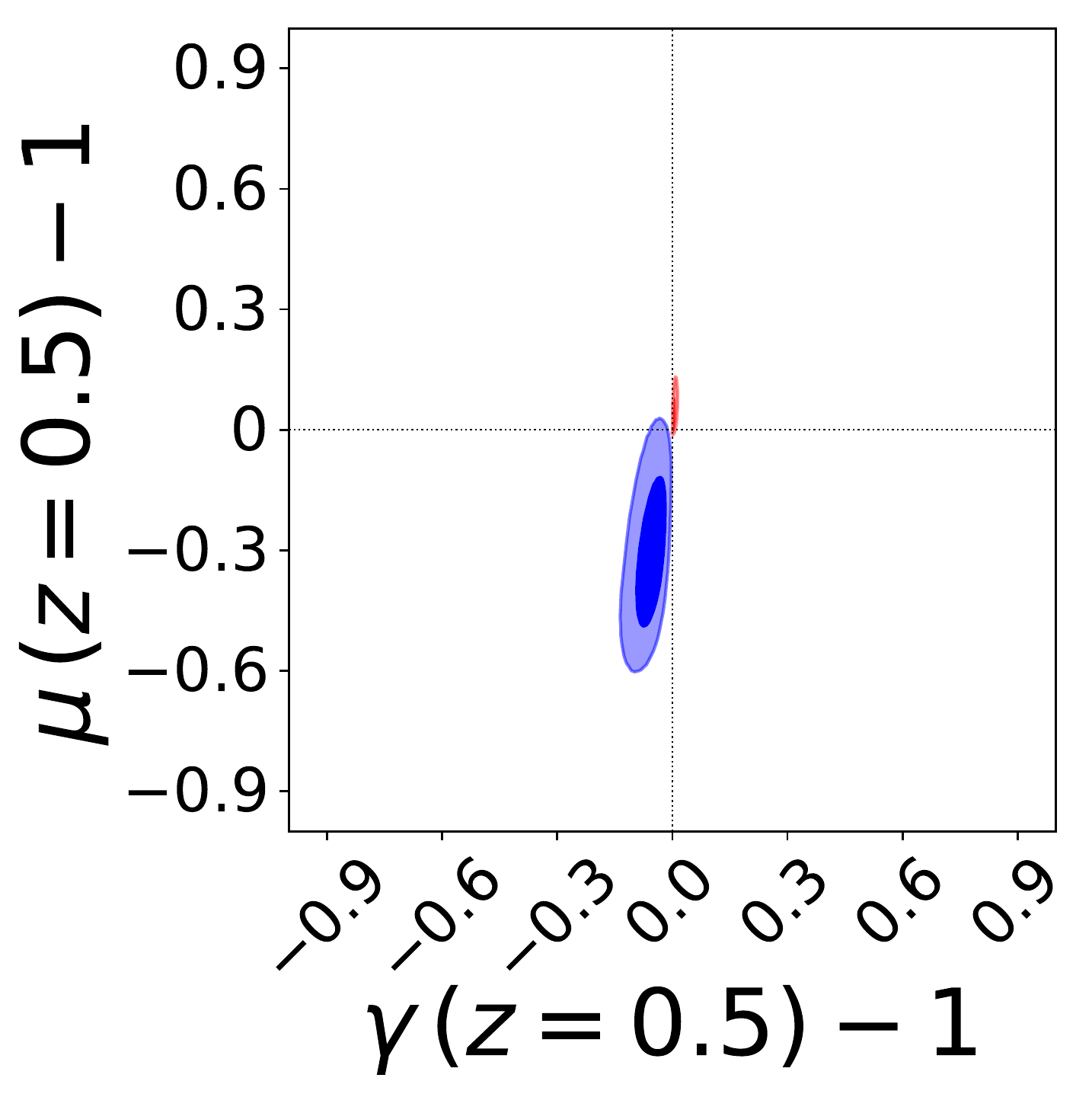}
\includegraphics[clip, trim = 0.05cm 0.05cm 0.05cm 0.05cm,scale=0.17]{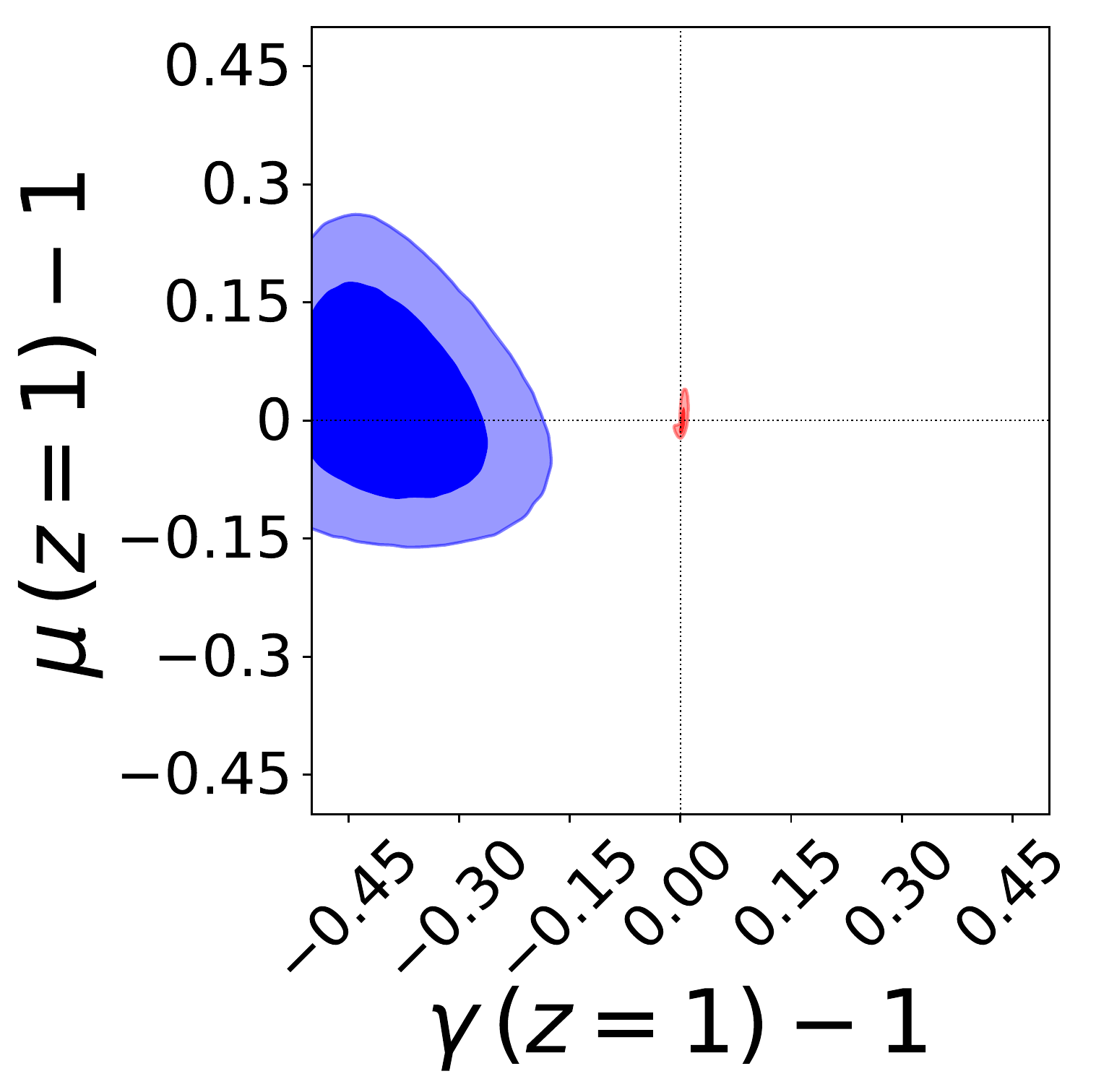}
\includegraphics[clip, trim = 0.05cm 0.05cm 0.05cm 0.05cm,scale=0.17]{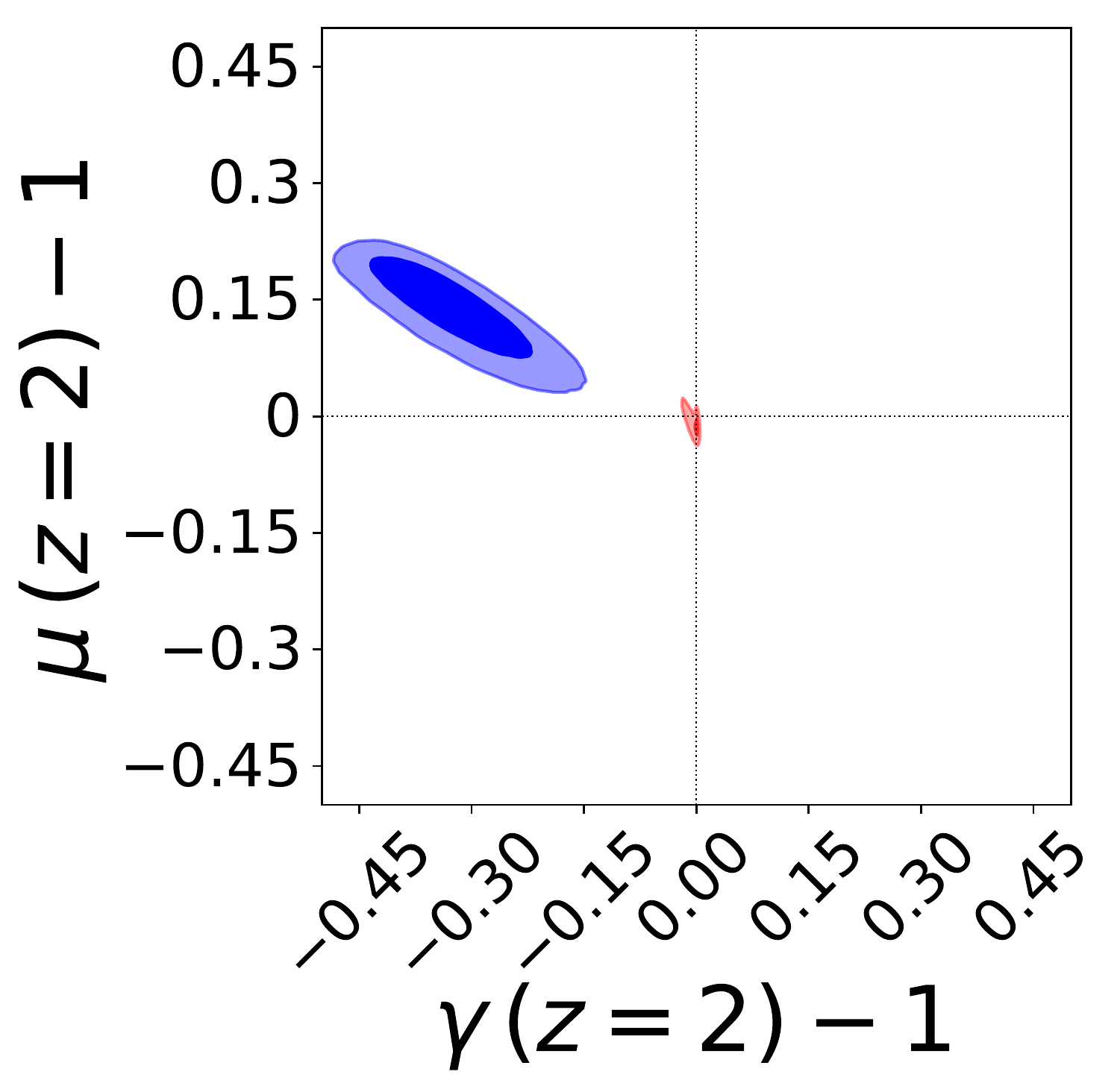}
\includegraphics[clip, trim = 0.05cm 0.05cm 0.05cm 0.05cm,scale=0.17]{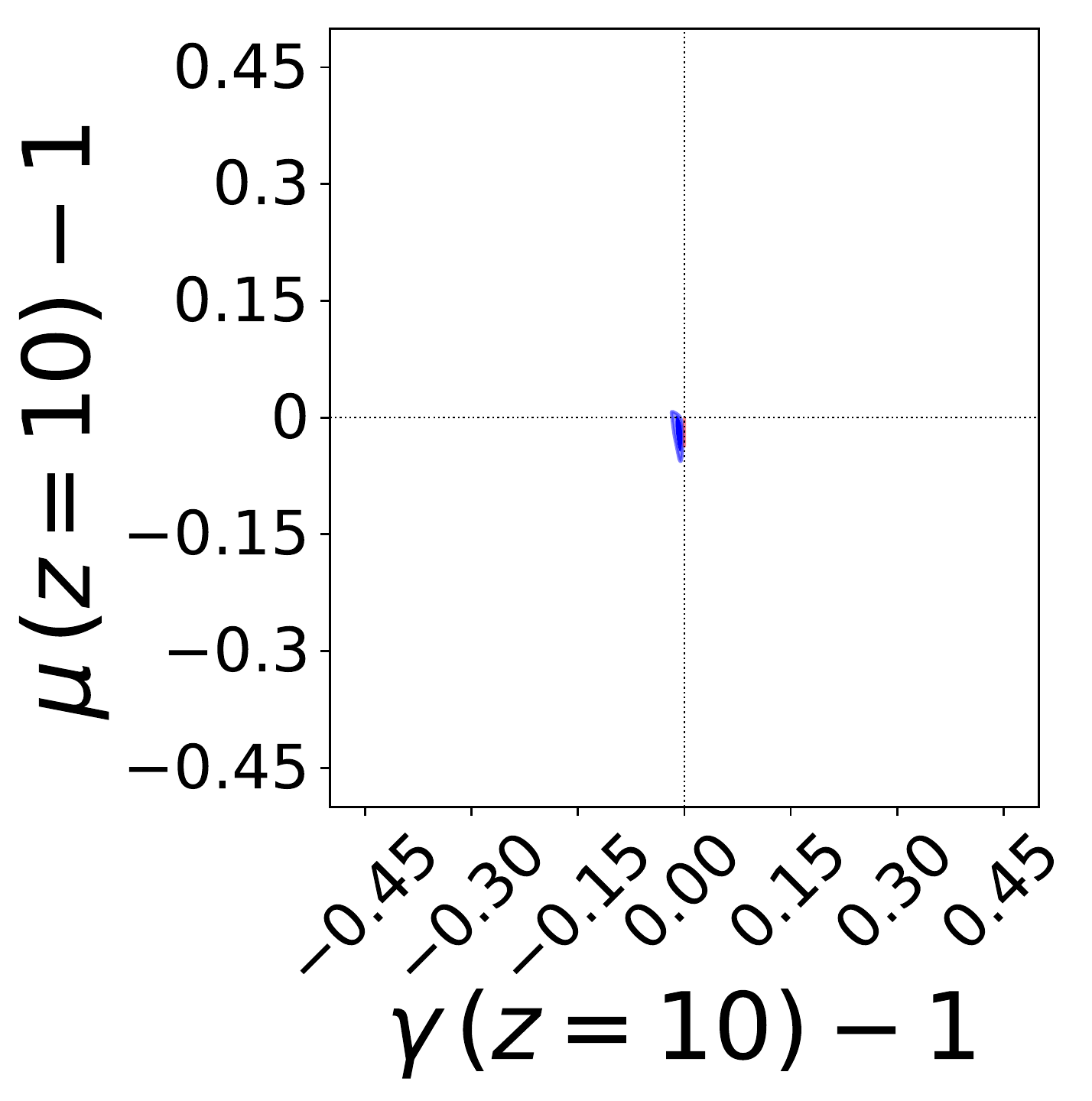}
\vskip2mm
\includegraphics[clip, trim = 0.05cm 0.05cm 0.05cm 0.05cm,scale=0.17]{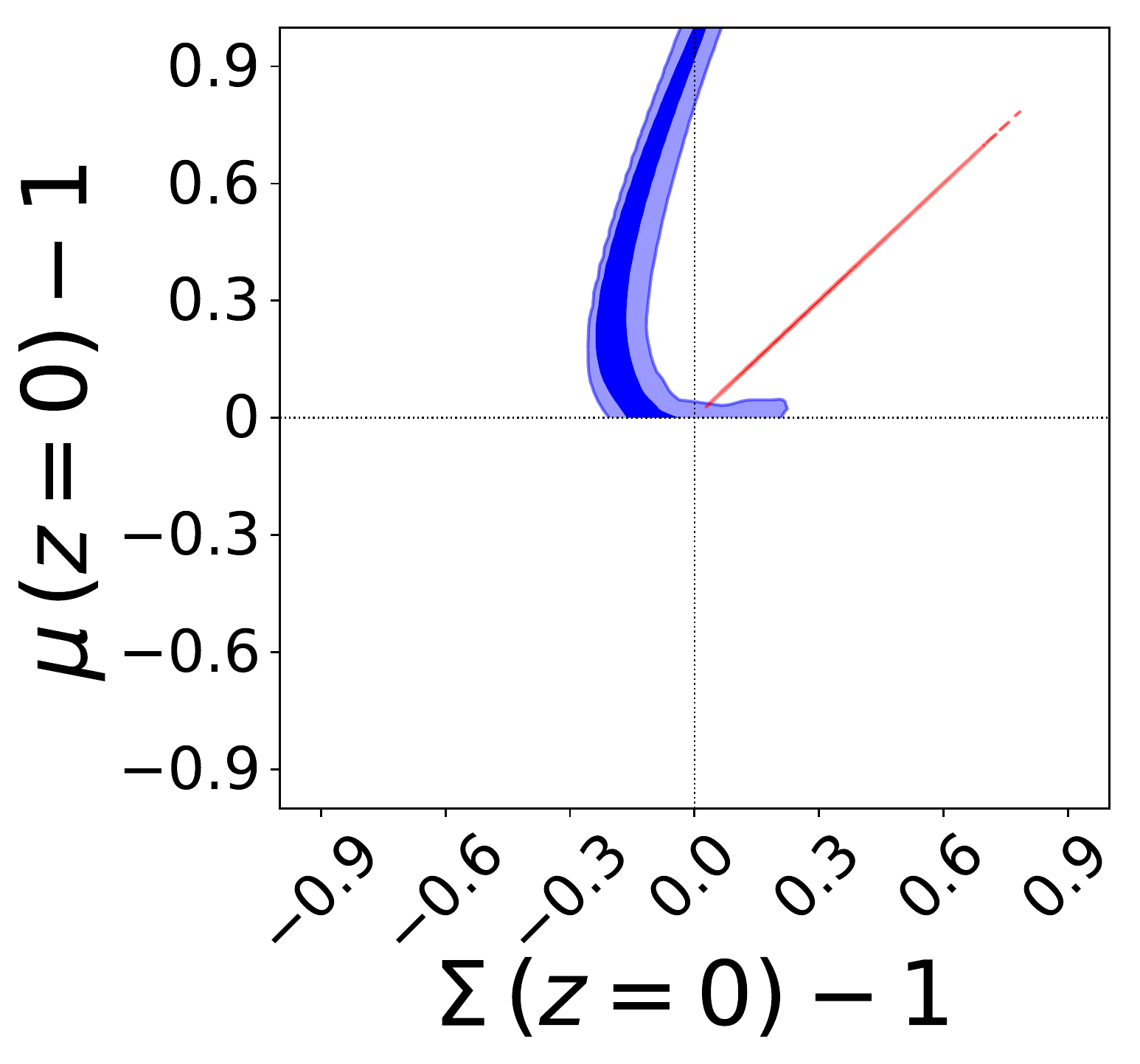}
\includegraphics[clip, trim = 0.05cm 0.05cm 0.05cm 0.05cm,scale=0.17]{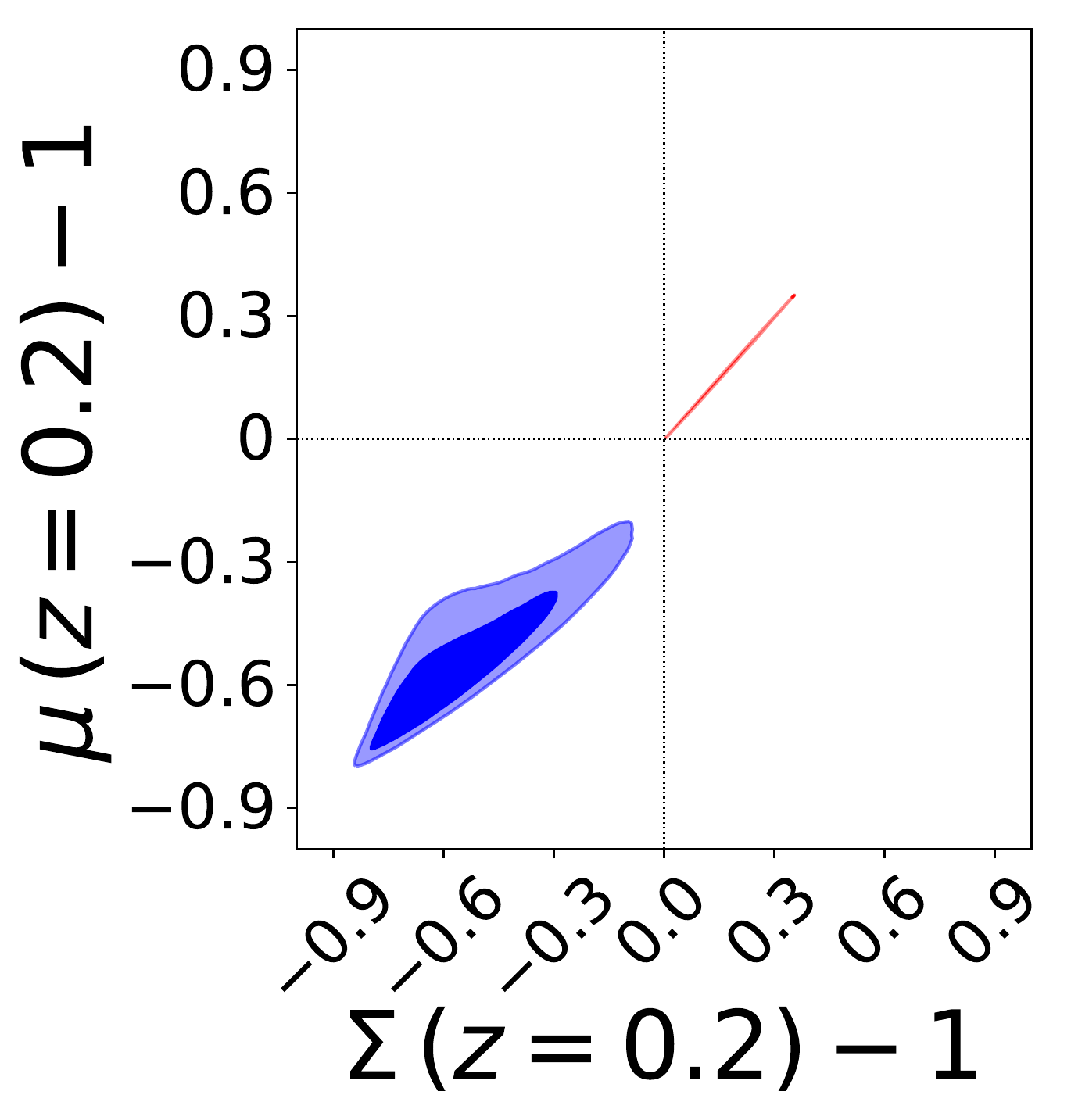}
\includegraphics[clip, trim = 0.05cm 0.05cm 0.05cm 0.05cm,scale=0.17]{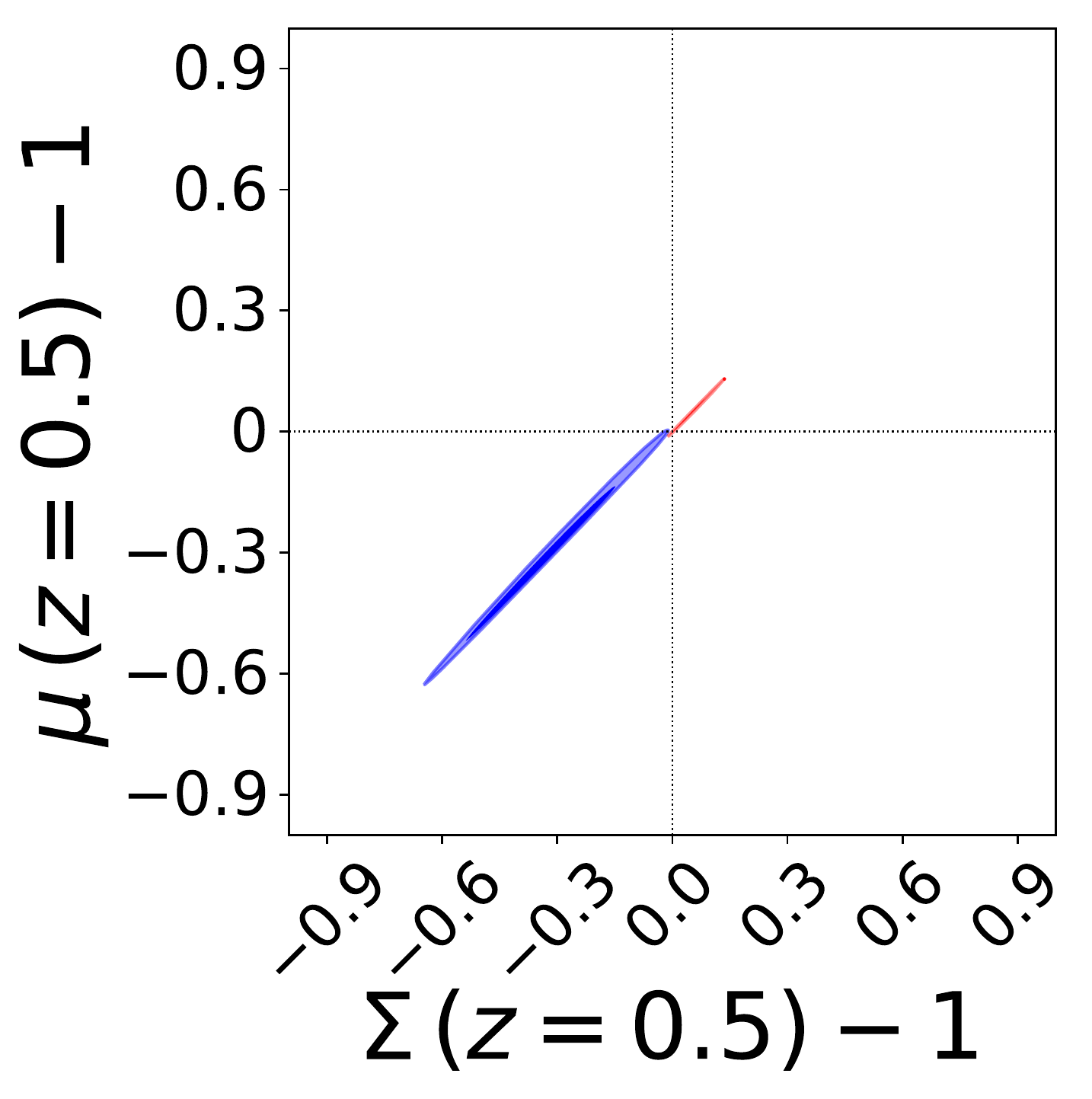}
\includegraphics[clip, trim = 0.05cm 0.05cm 0.05cm 0.05cm,scale=0.17]{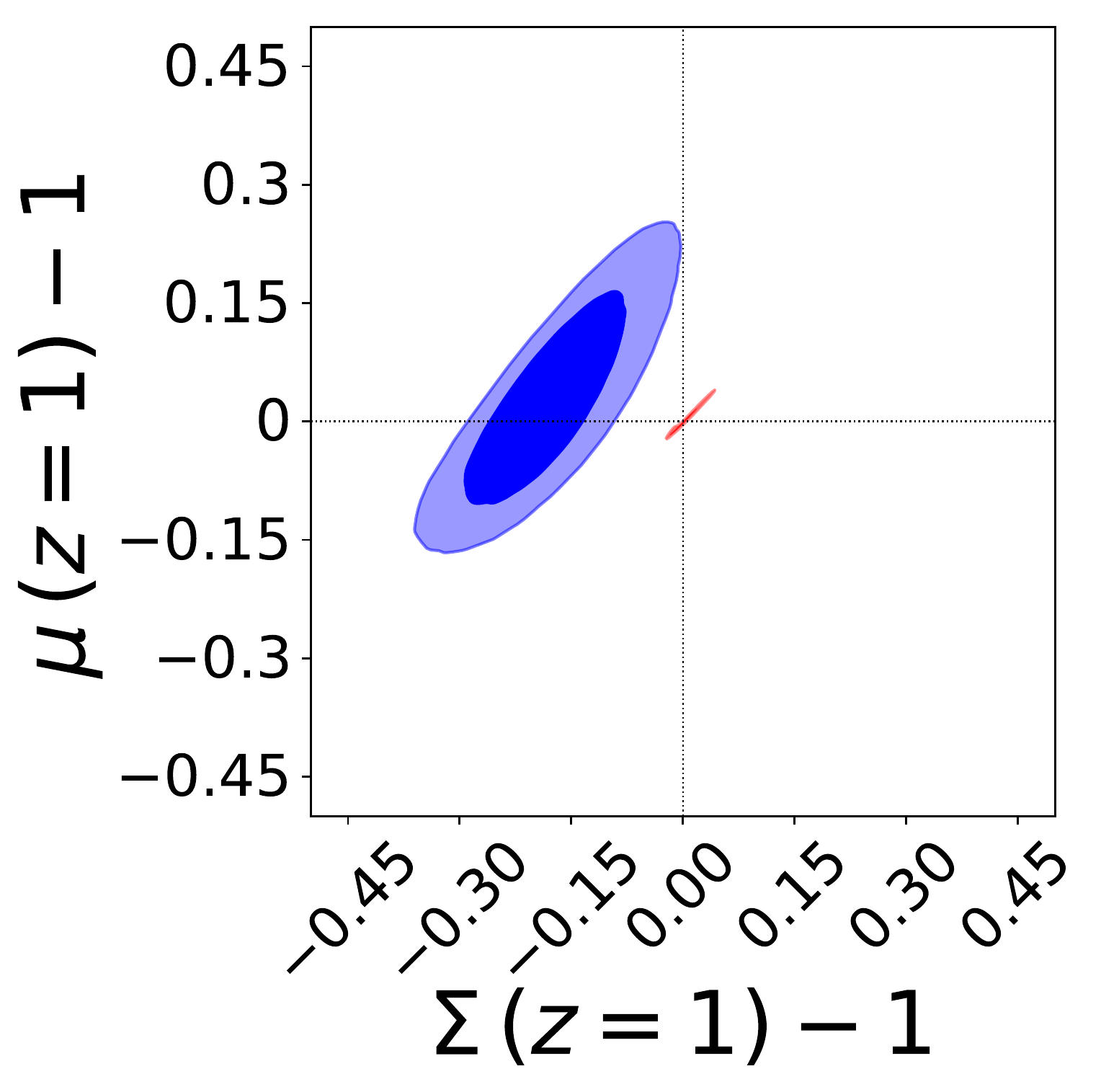}
\includegraphics[clip, trim = 0.05cm 0.05cm 0.05cm 0.05cm,scale=0.17]{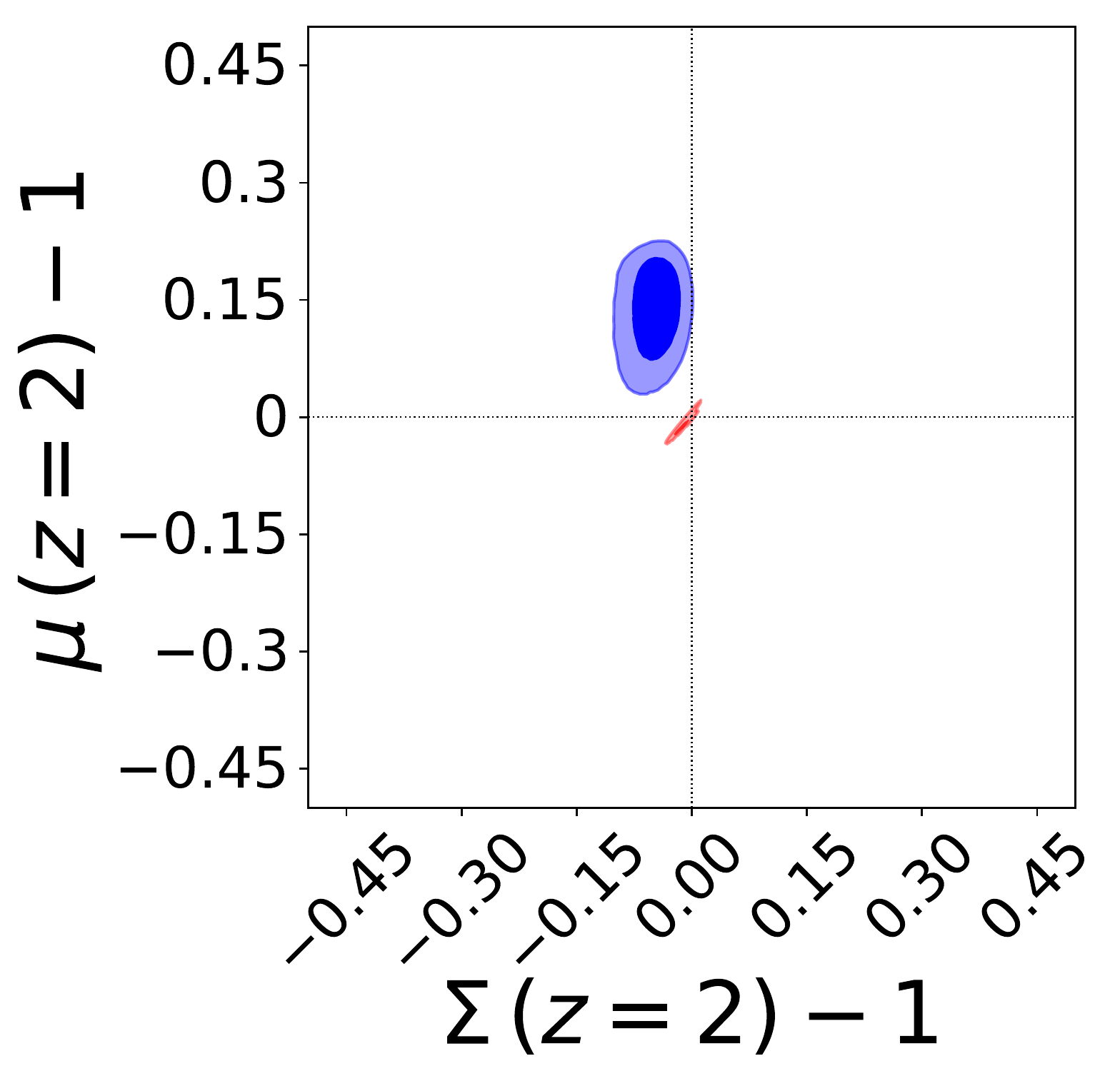}
\includegraphics[clip, trim = 0.05cm 0.05cm 0.05cm 0.05cm,scale=0.17]{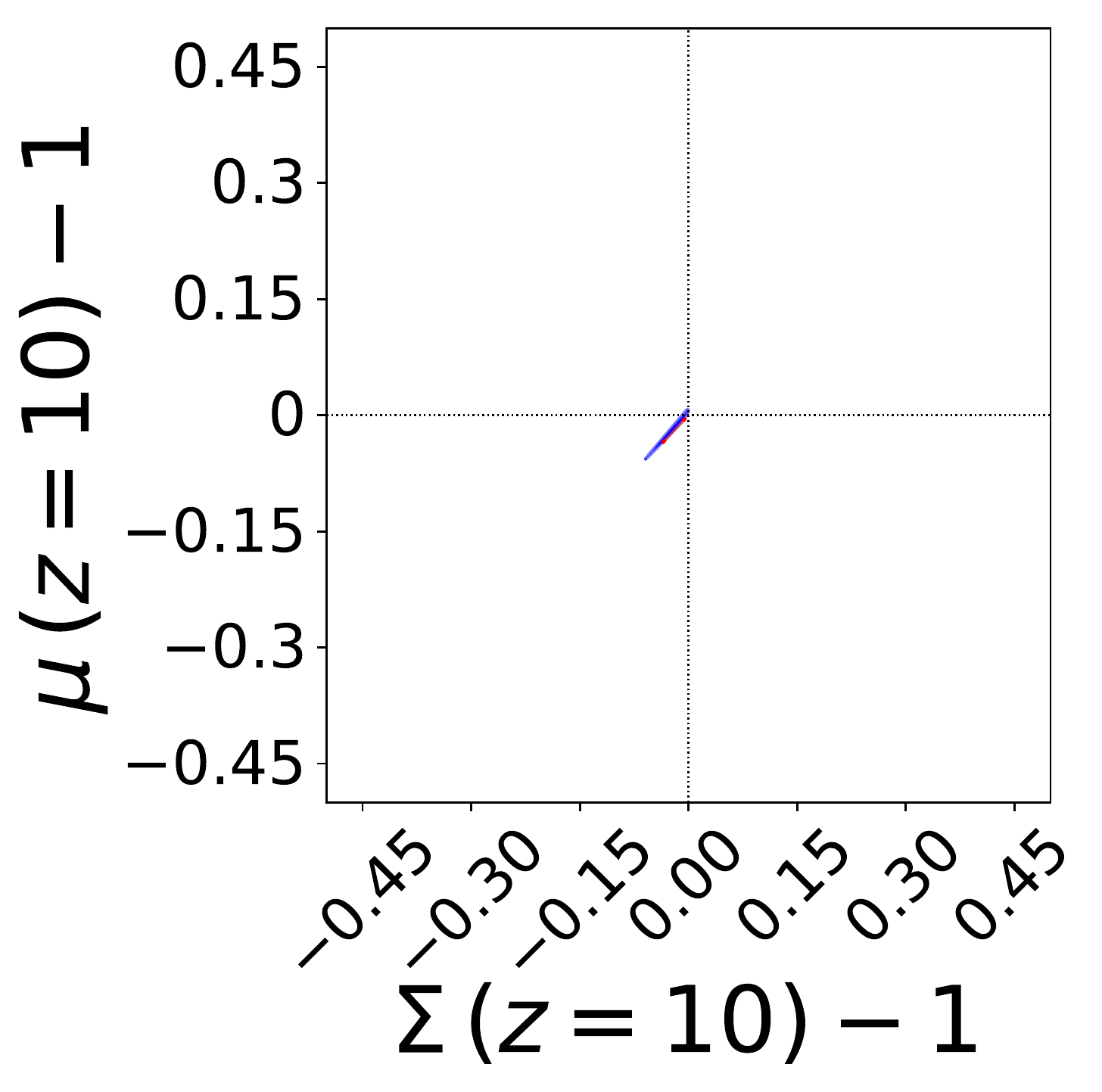}
   \caption{Redshift evolution of the 2D posterior distributions of the pairs $\lbrace \mu,\,\gsp \rbrace$ and $\lbrace \mu,\,\Sigma \rbrace$ from the constraints considering the \fspfps combination with and without the $\gdot$ prior. The axis scaling is not the same between the first three columns and last three columns.}
   \label{fig:obsredshift}
\end{center}
\end{figure}

A particularity of linear Horndeski theories is that by definition $\mu(z=0)\geqslant 1$ given the normalisation $M^2(z=0)=\mps$, $i.e.$ $\musc(z=0)=1$. Since stability implies $\muff(\forall z)\geqslant 1$, viable predictions can only lie in the $\mu-1>0$ upper quadrant at redshift 0, as already seen in Section \ref{sec:obs1}. The second particularity of Horndeski theories as noted in \cite{Perenon:2015sla,Perenon:2016blf} is the alternation of weaker/stronger gravity pattern in the LSS functions across redshifts. This is apparent in Figure \ref{fig:obsredshift} where the marginalised posteriors of the LSS functions obtained from the data constraints are displayed for several redshift bins. After $z=0$, at low redshifts the data favours weaker gravity $\lbrace \mu<1,\,\Sigma<1 \rbrace $ where the $\musc$ contribution dominates over $\muff$, at intermediate redshifts the opposite happens where $\muff$ dominates over $\musc$ and stronger gravity $\lbrace \mu>1,\,\Sigma>1 \rbrace $ is favoured. At higher redshifts, the couplings switching off as $1-\om$, causing the fifth force contribution and $\gsp$ to do so as well. The period of weaker gravity $\lbrace \mu<1,\,\Sigma<1, \gsp \sim 1 \rbrace $ arises again however, constrained further by the Planck prior as the redshift increases. 

\renewcommand{\arraystretch}{1.4}
\begin{table}[!]
\begin{center}
\begin{tabular}{|c||c|c|c|c|c|c|c|c|}
\hline 
68\% limits                               & $p_{10}$  & $p_{11}$ & $p_{20}$ & $p_{21}$ & $p_{30}$ & $p_{31}$  \\ 
\hline\hline 
$\fs+f+\sig$                              & $-6.643^{+3.686}_{-3.834}$    & $23.888^{+13.998}_{-13.403}$   & $-4.960^{+18.954}_{-20.924}$   & $26.215^{+33.959}_{-37.211}$   & $1.746^{3.699}_{-3.655}$   & $13.838 ^{12.946 }_{-12.241}$  \\
\hline 
$\fs+f+\sig+\gdot$                        & $-0.000^{+0.002}_{-0.002}$  & $-0.127^{+0.095}_{-0.096}$   & $1.697^{+2.933}_{-2.157}$   & $-0.926^{+5.852}_{-5.990}$      & $1.022^{+0.930}_{-0.806}$   & $-1.447^{+1.510}_{-1.812 }$  \\
\hline	
\end{tabular} 
\vskip2mm
\begin{tabular}{|c||c|c|c|c|c|c|c|c|}
\hline 
                                          & $\omo$  & $\so$ & $\mu\,(z=0)$ & $\Sigma\,(z=0)$ & $\gsp\,(z=0)$   \\ 
\hline\hline 
$\fs+f+\sig$                              & $0.320^{+0.026}_{-0.026}$ & $0.789^{+0.042}_{-0.046}$ & $1.575^{+1.021}_{-0.575}$ & $0.925^{+0.344}_{-0.165}$ & $0.240^{+0.615}_{-0.342}$ \\
\hline 
$\fs+f+\sig+\gdot$                        & $0.303^{+0.024}_{-0.024 }$ & $0.776^{+0.036}_{-0.037}$  & $1.321^{+0.370}_{-0.282}$ & $1.321^{+0.371}_{-0.280}$ & $1.00000\pm 0.00093$  \\	
\hline
\end{tabular} 
\caption{Mean values and 95\% confidence limits on the free parameters and LSS observables at redshift zero from the constraints with the \fspfps combination with and without the $\gdot$ prior.}
   \label{tab:paramlimits_gdotg}
\end{center}
\end{table}

The $\gdot$ prior has a drastic effect on the observables and alters the previous diagnostic. As said previously, this prior stringently bounds $\musc$ which enhances the effects of $\muff$. In doing so the alternation of weaker/stronger gravity is reduced to two periods, the stronger gravity $\lbrace \mu>1,\,\Sigma>1 \rbrace $ at low redshifts and at intermediate redshifts the weaker gravity $\lbrace \mu<1,\,\Sigma<1 \rbrace $ occurs and remains. 

\vv
The most striking effect of this prior is to strongly hold $\gsp$ near one across redshifts, implying the correlation $\mu$ - $\Sigma$ is forced to follow the $45^{\circ}$ line\footnote{A sub-class of models within Horndeski theories exhibiting no gravitational slip were found and studied in \cite{Linder:2018jil,Denissenya:2018mqs} and constrained with CMB data in \cite{Brush:2018dhg}. There, the models dubbed ``No Slip Gravity" were analytically engineered to exhibit $\gsp=1$ across all times. This was achieved by computing the analytical relations the coupling functions must satisfy to ensure $\gsp=1$ and such models also have $c_T=1$. Here, our models have their gravitational slip forced close to unity by observational bounds. Yet the results between the studies coincide well. Notably, they both highlight models favouring suppressed growth relative to the standard model.}. Obtaining $\gsp$ close to unity was expected analytically since this prior sets a stringent bound on the coupling $M^2$ and $\mu_1$ at low redshifts. Suppose for the sake of simplicity that $\mu_1 \approx 0$. Then, the expression of the gravitational slip parameter eq. \eqref{gsp} implies directly $\gsp \approx 1$ which, in turn, induces from \eqref{Sigma} that $\mu -1 \approx \Sigma-1$. The strong likeliness of viable Horndeski theories to produce $\mu$ and $\Sigma$ correlated was shown in \cite{Espejo:2018hxa}. The generation of a vast sample of healthy models and the inclusion of weak observational priors, and notably on $M^2$, showed that the predictions on these two observables were found to exhibit a correlation very close to one across all redshifts tested, hence a $\gsp$ close to unity.

Accordingly, the results pair well with the sign conjecture postulated in \cite{Pogosian:2016pwr}, and further shown in \cite{Peirone:2017ywi}, which states that viable Horndeski theories should display a sign agreement between $\mu$ and $\Sigma$ at all redshifts. As explained in \cite{Pogosian:2016pwr,Peirone:2017ywi}, the crucial factor for the conjecture to hold stems from the competition of the bare Planck mass and the speed of tensor modes. Having set $c_T=1$ in this present analysis, we fall in the case where the conjecture is recovered via the bounding of the evolution of $M^2$. This can be achieved as in the previous section with the 1D parametrisation, or when more freedom is allowed by applying the $\gdot$ prior. 

\vv
In terms of stringent bounds, at redshift zero the $\gdot$ prior induces the combination of \fspfps to produce an astonishing constraint on the gravitational slip parameter (see Table \ref{tab:paramlimits_gdotg})
\begin{equation}
\gsp = 1.0000 \pm{9.3 \times 10^{-4}}  \;\;(95\%\;{\rm c.l.})\;.
\end{equation}
Table \ref{tab:paramlimits_gdotg} also shows that the growth of structure data favour values of $\mu$ today slightly larger than unity, $i.e.$ larger than in GR, whether the $\gdot$ prior is applied or not. The $\gdot$ prior pushes $\mu$ to even larger values due to the domination of $\muff$ over $\musc$. Having $\mu(z=0)>1$ inevitably means favouring a fifth force at present time: since by definition $\musc(z=0)=1$, the expression \eqref{eq:mu} implies that any constraints on $\mu-1$ today correspond directly to constraints on the fifth force contribution $\muff$ and we obtain the bound (see Table \ref{tab:paramlimits_gdotg}) of
\begin{equation}
\muff = 1.321^{+0.370}_{-0.284} \;\;(95\%\;{\rm c.l.})\;,
\end{equation}
signalling a fifth force contribution at more than two sigma. However, this must be thoroughly assessed in the future considering all the relevant cosmological probes. Nevertheless, it is clear from this analysis that \mod models in light of growth of structure constraints lead to stronger gravitational coupling today, $i.e.$ favouring a fifth force, and yet manage to produce a suppression of growth with respect to the standard model. This occurs despite the weaker gravity behaviour being savagely cut by the bound on the variation of the Newton constant, the ingredients being manufactured at high redshifts where the slightly weaker gravity behaviour remains whereas the fifth force vanishes.

\vv
To conclude the analysis of the constraints on the LSS functions we must discuss a caveat when comparing our results to model-independent constraints. Our results are obtained by mapping the constraints on the free parameters of our Horndeski models onto the LSS functions $\mu$, $\Sigma$, $\gsp$, using the expressions in Section \ref{sec:characLSS}. In the literature, these functions have also been constrained using an alternative  more model-independent approach. For instance, a direct and blind parametrisation of the form  \cite{Simpson:2012ra,Ade:2015rim,Ferte:2017bpf,Aghanim:2018eyx} $\mathcal{O}(t)=1+\mathcal{O}_0 \times \Omega_{\rm de}(t)$ for each functions can be considered. These do not depend on a given gravitational theory but deviations from the standard model are allowed by a constant $\mathcal{O}_0$ for each observable. Doing so, the Planck Collaboration has shown in \cite{Ade:2015rim,Aghanim:2018eyx} that the combination of all relevant cosmological probes can lead to deviations from $\Lambda$CDM of more than $2\sigma$ in the LSS observables. These constraints favour values of $\mu (z=0)<1$ in line with the preference of data for lower values of $\sigma_{8,0}$ than in the standard model, and  dwell thus in a region unstable for Horndeski theories since only $\mu(z=0)\geqslant 1$ is allowed. One can be tempted to conclude that should future and more precise data continue in this direction, Horndeski theories would be observationally ruled out. However, such conclusions must be treated with care. As we have said, the fact that $\mu(z=0) \geqslant 1$ is a consequence of having chosen to define the bare Planck mass $M^2(z=0)=\mps$. The model-independent phenomenological parametrisation cannot encapsulate such a definition, highlighting the difficulty in linking astrophysical considerations and cosmological observations when adopting a model-independent approach. As shown in \cite{Espejo:2018hxa,Frusciante:2018jzw}, if the normalisation $M^2(z=0)=\mps$ is not required Horndeski theories are likely to display $\mu (z=0)<1$.

It was shown (see for instance \cite{Perenon:2015sla}) that even simple scalar-tensor theories such as Brans-Dicke theories produce time evolutions of the LSS functions more involved than that of a straight line as a function of the dark energy density. Over-simplified model-independent parametrisations are hence likely to miss the signatures even when little freedom is allowed to model departures from the standard model. Furthermore, \cite{Song:2010fg} allowed for a scaling in the phenomenological model-independent parametrisations of the LSS functions, and, indeed, depending on its value the width of constraints can vary significantly (see Figure 1 in \cite{Song:2010fg}). One has also to keep in mind that a model-independent phenomenological parametrisation of $\mu$ and $\Sigma$ cannot retain a link between the two. They are, in fact, not independent in a given theory and are thus bound to be correlated. Imposing a correlation between the two is likely to change the shape the model-independent constraints yield. For fairer confrontations with modified gravity predictions, we emphasise therefore the need to asses more involved model-independent parametrisations of the LSS functions in view of future surveys. 

\section{Conclusion}

We assessed in this paper the capacity of growth of structure data to constrain departures from the standard model induced by Horndeski theories with the speed of gravitational waves fixed to that of light. Using the effective field theory of dark energy we showed how the inclusion of the growth rate $f$ and the amplitude matter fluctuations $\sig$ measurements from VIPERS and SDSS, in addition to the standard redshift-space distortion data, significantly improves the constraints on the parameters of the model, by at least 20\%. Constraints on the  effective gravitational coupling $\mu$, gravitational slip parameter $\gsp$ and light deflections parameter $\Sigma$ were also obtained. We pointed out the importance of the generality of the parametrisation of linear Horndeski theories when constraining observables predictions. We also discussed the link with model-independent results of the literature and underlying caveats. 

The splitting of the effective gravitational constant into two distinct contributions was instrumental in understanding the behaviour of Horndeski theories in light of growth of structure constraints. The un-screenable coupling $\musc$, or equivalently the inverse of the bare Planck mass, is responsible for the suppression of growth across matter domination, while the fifth force counter part $\muff$ strengthens gravitational interactions at low redshifts. This splitting also allowed us to apply Solar System  tests, which stringently bound the variation of the Newton constant, as a prior on the unscreenable contribution to gravity. With the application of this prior, we found the constraining power of $f$ and $\sig$ measurements to be increased tenfold. The sensitivity of the combination of the two on both gravity contributions of the gravitational coupling can then be more clearly disentangled and the precision on the parameters when the prior is included with the redshift-space distortion data is at least quadrupled. 

\vv
The combination of all the growth of structure data and the Solar System  prior has led us to a number of generic conclusions. Supplemented by the normalisation of matter fluctuations at high redshift in accordance to Cosmic Microwave Background constraints, we found \emph{the evolution of the gravitational slip parameter to be strongly bounded close to unity thereby enforcing the effective gravitational coupling and light deflection parameter to be practically equal to each other across matter domination}. We showed that in linear Horndeski theories upon appropriately normalising the present value of the bare Planck mass, the screened contribution to the effective gravitational constant is forced to be unity at present time. This leads to \emph{the present value of the gravitational slip parameter being constrained to unity at the per thousand level}. The same characteristic of Horndeski theories also implies the present value of the gravitational constant can only be larger than one by definition and therefore any deviations must be attributed to the fifth force contribution. From this, we observe \emph{the growth of structure data to favour a fifth force contribution to the effective gravitational coupling at low redshifts and at more than two sigma at present time.} The latter might seem contradictory with the suppression of growth relative to the standard model. However, the fifth force is manifest at low redshifts whereas the slightly weaker gravity behaviour induced by the bare Planck mass is effective throughout matter domination. In fact, we find that \emph{despite abruptly cutting the predictions of weaker gravity in Horndeski theories, the inclusion of the Solar System  bound on the variation of the bare Planck mass does not prevent the production of suppressed growth with respect to the standard model}. The restriction of the space of viable models, induced by the coupling of theoretical and observational priors to the growth of structure data, allowed us to also constrain the redshift evolution of the one coupling which takes part only in the stability conditions when the quasi-static approximation is considered. However, we found these constraints to dilute significantly when mapped into the kineticity coupling $\ak$ of the $\alpha$-basis description of linear Horndeski theories.  

\vv
While the current measurements of $f$ and $\sig$ are in tension with the standard model, upcoming surveys of the large scale structure might well deepen the tension or consolidate the standard model. Nevertheless, obtaining more of these data separated will greatly enhance the power of growth of structures to probe departures from standard gravity. Improvements on the precision will enable the breaking of dark degeneracies in modified gravity in terms of fundamental parameters of the models, or on the physical phenomena behind observables predictions. 

\section*{Acknowledgements}

We warmly thank Jose Beltr\'an Jim\'enez, Guadalupe Ca\~nas Herrera, Sylvain De la Torre, Noemi Frusciante, Stephane Ili\'c, Eric Jullo, Christian Marinoni, Matteo Martinelli, Simone Peirone, Levon Pogosian and Alessandra Silvestri for fruitful discussions during this work. Numerical computations were done on the Sciama High Performance Compute Cluster which is supported by the ICG at the University of Portsmouth. LP thanks Sulona Kandhai and Miguel M\'endez Isla for their help and gives many thanks to Gary Burton for the support through the quirks of numerical endeavours. LP acknowledges financial support from the National Research Foundation (South Africa) (Grant Nos. 110966 and 119652). RM is supported by the South African Radio Astronomy Observatory and the National Research Foundation (Grant No. 75415),  and by the UK Science \& Technology Facilities Council (Grant No. ST/N000668/1). AdlCD acknowledges financial support from FIS2014-52837-P (Spanish MINECO), FIS2016-78859-P (AEI / FEDER, UE,  European Regional Development Fund and Spanish Research Agency (AEI), Consolider-Ingenio MULTIDARK CSD2009-00064 (Spanish MINECO), CA16104 and CA15117 CANTATA COST Actions EU Framework Programme Horizon 2020, CSIC I-LINK1019 Project, Spanish Ministry of Economy and Science, the University of Cape Town Launching Grants Programme and the National Research Foundation (South Africa) (Grant Nos. 99077 and 110966).

\bibliographystyle{JHEP}
\bibliography{References.bib}
\end{document}